\input harvmac 
\input epsf.tex
\def\IN{\relax{\rm I\kern-.18em N}} 
\def\IR{
\relax{\rm I\kern-.18em R}} \font\cmss=cmss10 
\font\cmsss=cmss10 at 7pt \def\IZ{\relax\ifmmode\mathchoice 
{\hbox{\cmss Z\kern-.4em Z}}{\hbox{\cmss Z\kern-.4em Z}} 
{\lower.9pt\hbox{\cmsss Z\kern-.4em Z}} {\lower1.2pt\hbox{
\cmsss Z\kern-.4em Z}}
\else{\cmss Z\kern-.4em Z}\fi} 

\overfullrule=0mm
\def\file#1{#1}
\newcount\figno \figno=0
\newcount\figtotno      
\figtotno=0
\newdimen\captionindent 
\captionindent=1cm 
\def\figbox#1#2{\epsfxsize=#1\vcenter{
\epsfbox{\file{#2}}}} 
\newcount\figno
\figno=0
\def\fig#1#2#3{ \par\begingroup\parindent=0pt
\leftskip=1cm\rightskip=1cm\parindent =0pt 
\baselineskip=11pt
\global\advance\figno by 1
\midinsert
\epsfxsize=#3
\centerline{\epsfbox{#2}}
\vskip 12pt
{\bf Fig. \the\figno:} #1\par
\endinsert\endgroup\par
}
\def\figlabel#1{\xdef#1{\the\figno}} 
\def\encadremath#1{\vbox{\hrule\hbox{\vrule\kern8pt 
\vbox{\kern8pt \hbox{$\displaystyle #1$}\kern8pt} 
\kern8pt\vrule}\hrule}} \def\enca#1{\vbox{\hrule\hbox{
\vrule\kern8pt\vbox{\kern8pt \hbox{$\displaystyle #1$}
\kern8pt} \kern8pt\vrule}\hrule}}

\def\IR{\relax{\rm I\kern-.18em R}}
\font\cmss=cmss10 \font\cmsss=cmss10 at 7pt 
\def\IZ{\relax\ifmmode\mathchoice
{\hbox{\cmss Z\kern-.4em Z}}{\hbox{\cmss Z\kern-.4em Z}} 
{\lower.9pt\hbox{\cmsss Z\kern-.4em Z}}
{\lower1.2pt\hbox{\cmsss Z\kern-.4em Z}} 
\else{\cmss Z\kern-.4em Z}\fi} \def\buildrel#1\under#2{ 
\mathrel{\mathop{\kern0pt #2}\limits_{#1}}}

\Title{T98/070}
{{\vbox {
\centerline{New Integrable Lattice Models}
\medskip
\centerline{From Fuss-Catalan Algebras}}}}

\bigskip
\centerline{P. Di Francesco\footnote*{e-mail: philippe@math.unc.edu},}
\bigskip
\centerline{\it Department of Mathematics,} 
\centerline{\it University of North Carolina at Chapel Hill,} 
\centerline{\it  CHAPEL HILL, N.C. 27599-3250, U.S.A.,} 
\medskip
\centerline{and}
\medskip
\centerline{\it Service de Physique Th\'eorique,}
\centerline{\it C.E.A. Saclay,}
\centerline{\it F-91191 Gif-sur-Yvette Cedex, FRANCE.}
\vskip .5in
\noindent  We construct new hyperbolic solutions of 
the Yang-Baxter equation, 
using the Fuss-Catalan algebras, a set of multi-colored versions of the 
Temperley-Lieb algebra, recently introduced by Bisch and Jones.
These lead to new two-dimensional integrable lattice models, describing
dense gases of colored loops.  
\Date{07/98}

\nref\BAX{R.J. Baxter, {\it Exactly Solved Models in Statistical
Mechanics}, Academic Press, London (1982).}
\nref\YJ{M. Jimbo, {\it Introduction to the Yang-Baxter equation},
Int. Jour. Mod. Phys. {\bf A4} No.15 (1989) 3759-3777.}
\nref\TRIGO{G. Delius, M. Gould and Y.-Z. Zhang, {\it On the construction
of trigonometric solutions of the Yang-Baxter equation}, Nucl.
Phys. {\bf B432} (1994) 337-403.}
\nref\RIMS{M. Jimbo and T. Miwa, {\it Algebraic analysis of 
solvable lattice models}, Regional 
Conference Series in Mathematics, Vol.85, A.M.S. (1994).}
\nref\TLA{H. Temperley and E. Lieb, {\it Relations between the Percolation
and Coloring Problems and other Graph-Theoretical Problems associated with
regular Planar Lattices: Some Exact Results for the Percolation
Problem}, Proc. Roy. Soc. {\bf A322} (1971) 251-280.}
\nref\MAR{P. Martin, {\it Potts Models and Related Problems in Statistical
Mechanics}, World Scientific, Singapore (1991).}
\nref\KAUF{L. Kauffman, {\it State models and the Jones polynomial}, 
Topology {\bf 26} (1987) 395-407.}
\nref\DGG{P. Di Francesco, O. Golinelli and E. Guitter,
{\it Meander, Folding
and Arch Statistics}, Mathl. Comput. Modelling,
Vol. {\bf 26}, No.8-10 (1997) 97-147.}
\nref\TLM{P. Di Francesco, O. Golinelli and E. Guitter,
{\it Meanders and the Temperley-Lieb Algebra},
Commun. Math. Phys. {\bf 186} (1997), 1-59.}
\nref\FC{D. Bisch and V. Jones, {\it Algebras associated to
intermediate subfactors}, Inv. Math. {\bf 128} (1997) 89-157.}
\nref\JA{M. Jimbo, T. Miwa and M. Okado, {\it An $A_{n-1}^{(1)}$
family of solvable lattice models}, Mod. Phys. Lett. {\bf B1} (1987)
73-79, and {\it Solvable lattice models related to the vector
representation of classical simple Lie algebras}, Comm. Math.
Phys. {\bf 116} (1988) 507-525.}
\nref\PAS{V. Pasquier, Nucl. Phys. {\bf B285} (1987) 162-172; J. Phys. 
{\bf A}:Math.Gen. {\bf 20} (1987) L1229, 5707.}
\nref\DIL{P. Roche, Phys. Lett. {\bf B285} (1992) 49-53;
S. Warnaar and B. Nienhuis, {\it Solvable lattice models
labelled by Dynkin diagrams}, J. Phys. {\bf A26} (1993) 2301-2316}
\nref\PEA{D. O'Brien and P. Pearce, {\it Lattice realizations of 
unitary minimal modular invariant partition functions}, 
J. Phys. {\bf A28} (1995) 261.}
\nref\MDET{P. Di Francesco and E. Guitter, work in progress.}


\newsec{Introduction}
\par
The Yang-Baxter equation \BAX\ plays a central role in the definition 
of two-dimensional integrable lattice models. It is a sufficient 
condition on the matrix of Boltzmann weights of the statistical model
ensuring integrability. Each of its solutions
indeed leads to an infinite family of commuting transfer matrices,
which should simultaneously diagonalize in a Bethe-Ansatz-type basis.
Solutions with additive spectral parameters are of three types:
elliptic, trigonometric (or hyperbolic) and rational,
referring to the type of functions of the spectral parameter involved
in the definition of the Boltzmann weights, each type leading to the 
next by some limiting process.

In this paper, we concentrate on hyperbolic solutions. Among the 
many known constructions, the most systematic seems to be to 
associate a solution to any multiplicity-free tensor product of two 
representations of an affine quantum algebra ${\cal U}_q({\widehat 
G})$, $G$ any classical Lie algebra \YJ\-\TRIGO. The simplest example 
of this uses two spin-$1\over 2$ representations of 
${\cal U}_q({\widehat sl}_2)$, and leads to the XXZ quantum spin 
chain or the equivalent 6 Vertex model \RIMS.
This solution actually arose from Baxter's study of the equivalence 
between the 6 Vertex model and the critical $Q$-states Potts model, 
in which he found the matrix of Boltzmann weights to be of the form
$W_i(u)=1_i+a(u)U_i$, $u$ the spectral parameter, where $U_i$ form
a matrix representation of the Temperley-Lieb algebra $TL_n(\beta)$ 
\BAX\ \TLA, with $\beta=\sqrt{Q}$. 

The Temperley-Lieb algebra has a simple pictorial representation 
which makes it ideal for describing a dense loop model on the square 
lattice, with a fugacity $\beta$ per loop. 
This representation has also become a basic tool in the definition
of link polynomials \KAUF.
A relation with meanders \DGG, 
i.e. compact folding configurations of polymer chains has also been
found \TLM, using the same representation.

The Temperley-Lieb algebra has been recently generalized by Bisch and
Jones \FC, by introducing a multi-colored version of the above 
pictorial representation. The resulting algebras are called the
$k$-color Fuss-Catalan algebras, denoted by 
$FC_n(\alpha_1,\alpha_2,\ldots,\alpha_k)$.
In the particular case $k=2$, the authors have given an explicit 
presentation of $FC_n(\alpha,\beta)$ in terms of generators and 
relations.

In this paper, we use the $k$-color Fuss-Catalan algebras to construct 
new hyperbolic solutions to the Yang-Baxter equation. The 
corresponding statistical models describe multi-colored dense loops 
with specific fugacities.
 
\medskip
The paper is organized as follows. 
We start by recalling in Sect.2 some known facts about the 
one-projector hyperbolic solutions to the Yang-Baxter equation,
and their link with the Hecke algebra, of which the Temperley-Lieb
algebra is simply the particular quotient corresponding to the 
commutant of ${\cal U}_q({\widehat sl}_2)$.
In Sect.3, we solve the Yang-Baxter equation, looking for a
matrix of Boltzmann weights which is a linear combination of the 
generators of $FC_n(\alpha,\beta)$, with coefficients function of the 
spectral parameter.
This is possible only if $\alpha=\beta$ and leads to the definition of
a dense two-loop model on the square lattice.
In Sect.4, we generalize this to the case of an arbitrary $k$-color 
Fuss-Catalan algebra, for which we first give a presentation with 
generators and relations. 
We find a number of solutions, indexed by a set of
``spins" $r_i\in \{\pm 1\}$, $i=2,4,\ldots,k-2$, $r_i=r_{k-i}$. 
These lead to integrable 
multi-colored dense loop models, with specific fugacities.
Sect.5 is devoted to a discussion of our solutions and concludes with 
a list of open questions to be addressed.
\par

\newsec{Hyperbolic Solutions To The Yang-Baxter Equation}
\par
\subsec{Vertex Models And The Yang-Baxter Equation}
\par
The Yang-Baxter equation is a sufficient condition for ensuring
the existence of an infinite set of commuting transfer matrices for
a given two-dimensional statistical model. In this work, we will
consider only square lattice vertex models. 
Let us consider a rectangle $R$ containing $N\times M$ vertices and, 
say $E$ edges numbered $1,2,\ldots,E$. 
A configuration of
the model is a map $\sigma$ from this set of edges to 
${\cal T}^{\otimes E}$, 
$\cal T$ a finite ```target" set. The weight of such a configuration is
a product over all the vertices of $R$ of the vertex Boltzmann weights
\eqn\bwt{ w(i,j\vert k,m) = \figbox{1.5cm}{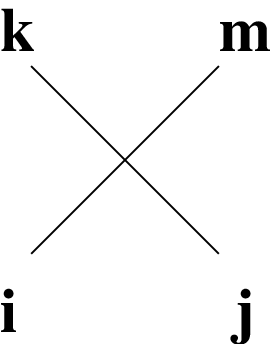} }
where $i,j,k,m$ denote the images of the four edges of the vertex.
The quantities of interest are statistical sums, such as
the partition function
\eqn\pfunc{ Z = \sum_{\rm configurations} \prod_{\rm vertices} 
w(i,j\vert k,m)  }

The target ${\cal T}$ being finite, it is useful to trade it for a vector 
space
$V$ of dimension $n=\vert {\cal T}\vert$, with a distinguished basis 
$\{e_1,\ldots,e_n\}$ in 
bijection with the elements of $\cal T$. We may then view the boltzmann 
weights \bwt\ as the matrix elements of a linear operator
$W: V\otimes V \to V\otimes V$, acting from top to bottom, namely
\eqn\actw{ W e_k \otimes e_m = \sum_{i,j} w(i,j\vert k,m) \ 
e_i \otimes e_j}

The operator $W$ may act on a line of $N+1$ edges 
$V\otimes V\otimes \ldots \otimes V$, as the identity on all spaces 
except on two of them, say in positions $(r,s)$, where it acts
as in \actw. We will denote $W_{r,s}$ the corresponding operator,
and  $W_i=W_{i,i+1}$. The monodromy matrix of the model is defined as
\eqn\mono{ {\cal M} = W_1 W_2 \ldots W_{N} }
and the transfer matrix $T$ is simply the trace of $\cal M$, 
viewed as a linear 
operator from $V_{N+1}$ to $V_1$.
It is now clear that, if we impose doubly periodic conditions along
the bordering edges of $R$, the resulting partition function reads
\eqn\pftr{Z={\rm Tr}(T^M) }
and that the diagonalization 
of $T$ allows for solving the model completely.

The Yang-Baxter equation is a sufficient condition for the existence of
an infinite set of commuting transfer matrices $T(x)$, $x$ a
real parameter entering the definition of the Boltzmann weights \bwt.
When the $T$'s are diagonalizable,this in turn grants the existence 
of a common basis of eigenvectors for all the $T(x)$, which can be
found by Bethe Ansatz techniques.
The Yang-Baxter equation reads:
\eqn\ybe{\encadremath{W_i(x) W_{i+1}(xy) W_i(y) 
= W_{i+1}(y) W_i(xy) W_{i+1}(x)}}
and is usually supplemented by the normalization condition:
\eqn\norma{\encadremath{W_i(x) W_i({1\over x}) = 1}}
which fixes the gauge $W_i(x) \to f(x) W_i(x)$ up to a factor $\rho(x)$
such that $\rho(x)\rho(1/x)=1$.
Note that both equations \ybe-\norma\ must hold for $i=1,2,\ldots,N$, 
whereas all operators act on a line of $N+1$ edges $V^{\otimes (N+1)}$
(in particular, $1$ stands for the identity $I\otimes\ldots\otimes I$).
It is customary to further fix the normalization of $W_i$ by imposing
\eqn\imposi{\encadremath{ W_i(1) = 1 }}

The solutions of the Yang-Baxter equation are known to be of three 
types: elliptic, trigonometric (or hyperbolic) and rational, 
referring to the type
of dependence on the spectral parameter $u$, such that 
$x=e^u$. Each type degenerates into the next in some limit.
Here we will mainly focus on hyperbolic solutions (i.e. involving
only rational fractions of $x$). 

\subsec{Hyperbolic Solutions: The One-projector Case}

Let us determine all solutions to \ybe\-\norma\ of the form
\eqn\projo{ W_i(x) = 1_i + a(x) U_i }
where all the dependence on $x$ is contained in the function $a(x)$, 
$1$ stands for the identity of $V\otimes V$, and $U$ is an endomorphism 
of $V\otimes V$.
The normalization condition \norma\ implies that
$U$ satisfies a quadratic equation of the form $\lambda U+\mu U^2=0$, 
and if $U$ is non-trivial, we must have $U^{2}= \beta U$. Hence $U$
is an un-normalized projector. Moreover, 
\eqn\noram{ a(x)+ a({1\over x}) +\beta a(x) a({1\over x}) =0} 
It is now easy to write the condition \ybe, which amounts to
\eqn\ybex{\eqalign{&\big( a(x)+a(y)+\beta a(x)a(y)
-a(xy) \big) (U_i-U_{i+1}) \cr
&\ \ \ \ \ \ \ \ 
+a(x)a(xy)a(y)( U_i U_{i+1} U_i- U_{i+1} U_i U_{i+1}) = 0\cr}}
Up to a multiplicative redefinition of $\beta$, this implies that
$ U_i U_{i+1} U_i- U_i= U_{i+1} U_i U_{i+1}-U_{i+1}$, and that
\eqn\aeq{ a(x)+a(y)+\beta a(x)a(y)+a(xy)\big(1-a(x)a(y)\big) =0}
Expanding to first order
in $\epsilon$, for $y=e^\epsilon/ x$, and using $a(1)=0$ from
\imposi, we find 
\eqn\difx{ {1\over x} a'({1\over x}) (1+\beta a(x))+ 
a'(1)(1-a(x)a({1\over x}) = 0}
which, together with \noram\ leads to the differential equation
$x a'(x)=-a'(1)(1+\beta a(x)+a(x)^2)$, easily solved as 
\eqn\sola{ a(x) = { x^\gamma -1\over z - x^\gamma/z} }
where $z+1/z=\beta$ and where $\gamma= a'(1)(z-1/z)$ can be safely set 
to $1$ (which amounts to redefining $x=e^{\gamma u}$).

We conclude that the most general non-trivial solution to 
\ybe\-\norma\-\imposi\
of the form \projo\ reads
\eqn\mostl{ W_i(x) = 1_i + { x-1 \over z- {x \over z}} U_i }
where the $U$'s satisfy the following relations
\eqn\hecke{\encadremath{
\eqalign{ U_i^2 &= \beta U_i \ \ i=1,2,\ldots,N\cr
U_i U_j &= U_j U_i \ \ \  {\rm for}\ \ \vert i-j\vert > 1 \cr
U_i U_{i+1} U_i- U_i &= U_{i+1} U_i U_{i+1}-U_{i+1} \ \ 
i=1,2,\ldots,N-1 \cr}}}
The algebra generated by the $U_i$, $i=1,\ldots,N$ and the identity 
$1$ is the Hecke algebra $H_{N+1}(\beta)$.

This exercise can be repeated in the case of more projectors, and 
higher algebras can be found. In this paper, we will present new 
solutions with arbitrary numbers of projectors.
 
It is also interesting to note that particular solutions corresponding
to quotients of the hecke algebra (obtained by imposing extra 
relations on the $U$'s) have been found. They correspond to the
$A_{n-1}$ models of \JA, related to the quantum groups 
$U_q({\widehat sl}_n)$. In the remainder of this section, we will 
concentrate
on the first of these quotients, also known as the Temperley-Lieb 
algebra.

\subsec{Temperley-Lieb Algebra, 6 Vertex, Potts And Dense Loop Models}

The Temperley-Lieb algebra $TL_{N+1}(\beta)$ is the quotient of 
the Hecke algebra $H_{N+1}(\beta)$ obtained by imposing the extra
relations 
\eqn\tla{ U_i U_{i+1} U_i = U_i \ \ {\rm and} \ \ U_{i+1}U_i 
U_{i+1} = U_{i+1} }
for $i=1,2,\ldots,N-1$.
The Temperley-Lieb vertex model, with Boltzmann weights \mostl\
involving the generators $U_i$ of $TL_{N+1}(\beta)$ can take many 
forms, depending on the choice of representation $(U,V)$.

Taking a two-dimensional representation $V=Vect(e_+,e_-)$, and 
the $4\times 4$ matrix of $U$ acting on 
$V\otimes V=Vect(e_+e_+,e_+e_-,e_-e_+,e_-e_-)$:
\eqn\sixv{ U = \pmatrix{0 & 0 & 0 & 0 \cr
0 & z & 1 & 0 \cr
0 & 1 & {1 \over z} & 0 \cr
0 & 0 & 0 & 0 \cr}}
we recover the celebrated 6 Vertex model,
with parameter $\Delta=-\beta/2$. 
Another representation is known to correspond to the critical
square lattice $Q$ states Potts model \BAX. It involves a $Q$-dimensional
representation $V$, and the relation to our parameter $\beta$ is
$Q=\beta^2$. 
The equivalence between 6 Vertex and Potts models
can be proved by noting that a slight modification of our 
Temperley-Lieb vertex model (essentially by boundary terms) 
makes its partition function independent of the particular
representation chosen \BAX.

A remarkable property of the Temperley-Lieb vertex model 
is that it can be reformulated
as a dense loop model on the square lattice as follows.

The Temperley-Lieb algebra $TL_{N+1}(\beta)$ has a pictorial 
representation in which each 
generator is a rectangular domino with an upper and a lower line
of points numbered $1,2,\ldots,N+1$, connected by pairs through 
non-intersecting lines (strings). 
We have the following generators:
\eqn\dominotla{1~=~\figbox{2.6cm}{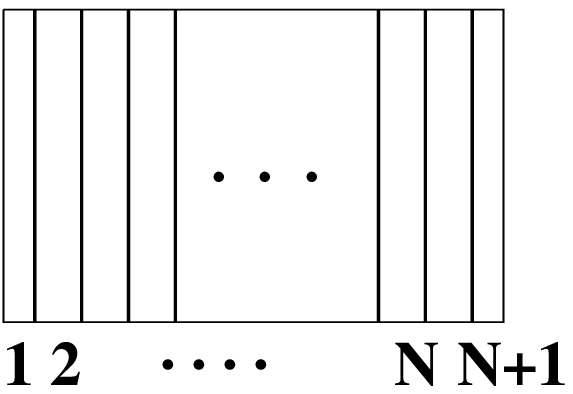} 
\qquad U_i~=~\figbox{2.6cm}{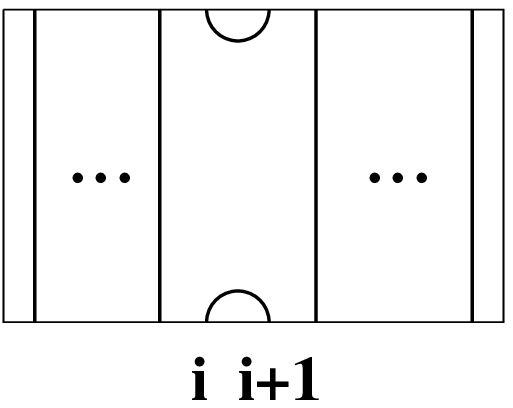}  }
A product of such generators is simply obtained by concatenation
of the corresponding dominos. The relations of $TL_N(\beta)$ become
transparent in the pictorial representation:
\eqn\tlrela{\eqalign{U_i^2~&=~
\figbox{2cm}{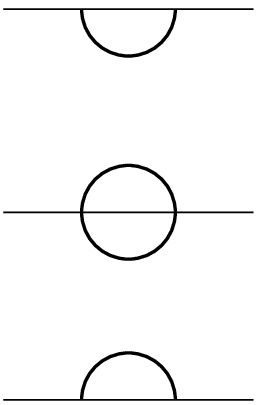}~=~q~\figbox{2.cm}{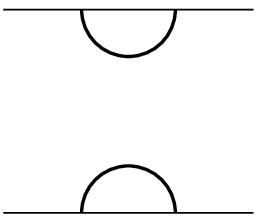}~=~q \, U_i\cr
U_i\, U_{i+1}\, U_i~&=~
\figbox{2.cm}{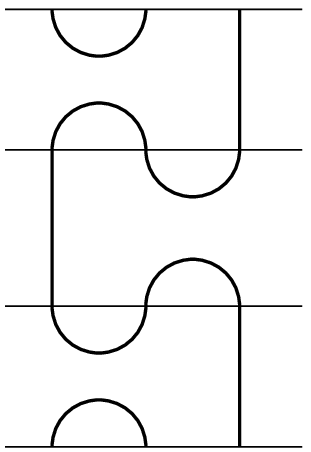}~=~\figbox{2.4cm}{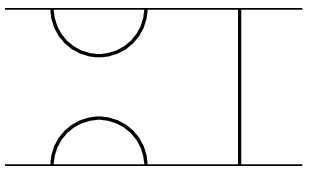}~=~ U_i\cr}}
In particular we see that strings may be pulled (second relation)
without altering the elements, and that loops may be replaced by a 
factor $\beta$ (first relation).
The most general elements of the algebra are linear combinations
of the reduced products of generators, namely products in which we
have pulled all strings and removed all loops. 

As the identity is represented with only vertical strings, we may
further simplify the notation and represent only the local part of
the generators which is not the identity: 
\eqn\simpu{U_i~=~ \figbox{1.5cm}{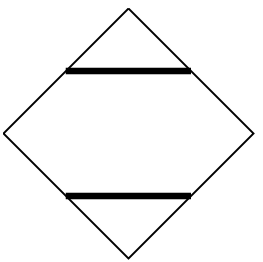}\ \ {\rm or} \ \
\figbox{1.5cm}{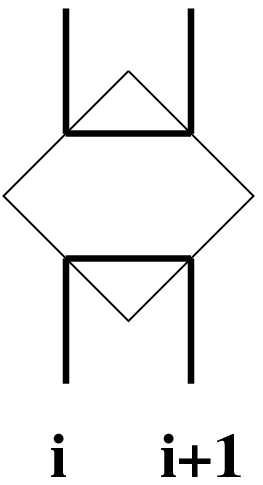} }
Another way of interpreting this is to view $U$ as an operator acting 
(say from top to bottom) on a pair of strings by creating a bridge. 
As illustrated in \simpu, $U_i$ corresponds to such
an action on two consecutive strings, with positions $i$ and $i+1$.
Analogously, we may represent the identity acting on the same strings 
as 
\eqn\identrep{1_i ~=~ \figbox{1.5cm}{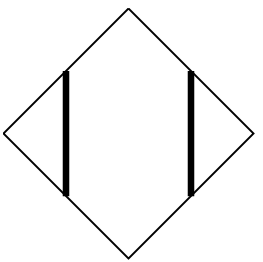}}

Consider now our Temperley-Lieb vertex model. The line of edges 
$V^{\otimes (N+1)}$ is now replaced by $N+1$ vertical strings.
The operator $W_i(x)$ is a linear combination
\eqn\wconf{W_i(x)~=~ \figbox{1.5cm}{ig.eps} + {x-1 \over z-{x\over z}}
\figbox{1.5cm}{ug.eps}}
where the strings have position $i$ and $i+1$. Imposing periodic 
boundary conditions along the top and
bottom ends of the strings, the partition function for a rectangle $R'$
containing $N\times M$ vertices reads (we assume $N$ and $M$ to be 
even)
\eqn\pfbc{Z~=~ {\rm Tr}\big( (T_e T_o)^{M/2}\big)}
where we have introduced even and odd transfer matrices
\eqn\eveno{\eqalign{ T_o &= W_1(x) W_3(x) \ldots W_{N-1}(x) \cr
T_e &= W_2(x) W_4(x) \ldots W_N(x) \cr}}
The trace used in \pfbc\ is the standard Markov trace of 
$TL_{N+1}(\beta)$,
defined on any reduced element by $\beta^k$, where $k$ is
the number of loops formed when we identify the top and bottom edges 
of the strings (e.g. Tr$(1)=\beta^{N+1}$, Tr$(U_i)=\beta^N$),
and is extended to any element of the algebra by linearity.
Comparing \pfbc\ with \pftr, we have in both cases
simply rotated our view of the
lattice by $45^{\circ}$, but the former rectangle ($R$) is rotated,
while the new one is not, namely 
\eqn\zcalc{ R=\figbox{3.5cm}{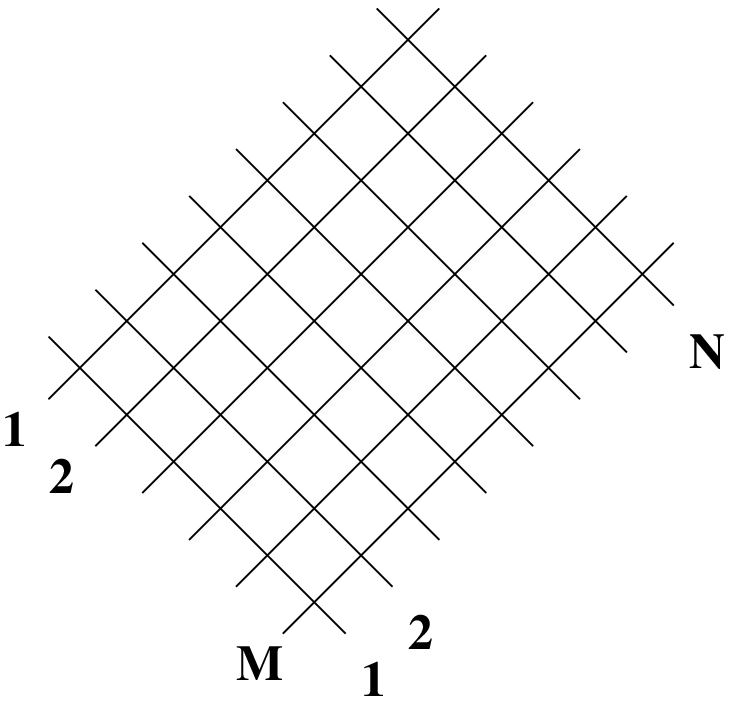} \ \ \ R'= \figbox{3.5cm}{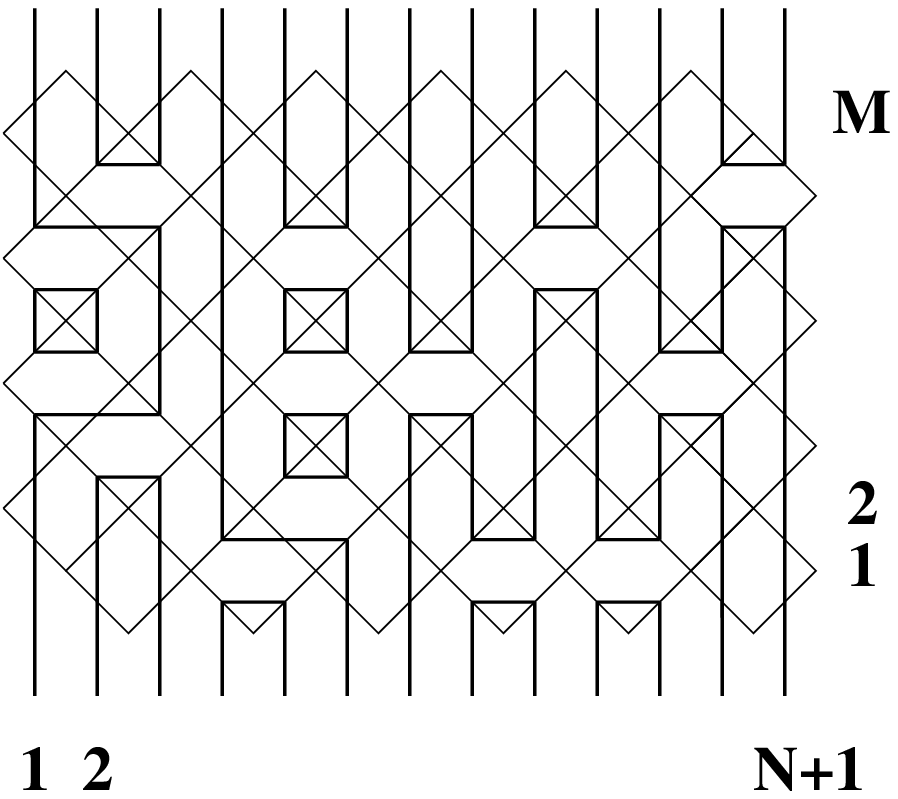}} 

The expression \pfbc\ suggests the following interpretation of the 
model. We consider the dual $R'^{*}$ of $R'$, in which each vertex 
becomes a square face, rotated by $45^\circ$. Expanding the transfer 
matrices in \pfbc\ by using the expression \wconf, we can rewrite
$Z$ as a sum over the $2^{NM}$ face configurations on $R'^*$ 
obtained by taking either the $1_i$ or the $U_i$ term in each 
$W_i(x)$, and we still have to take the trace of this sum in 
$TL_{N+1}(\beta)$:
\eqn\pfsum{Z~=~ \sum_{{\rm face}\ {\rm configurations}} 
a(x)^k {\rm Tr}(U_{i_1} U_{i_2} \ldots U_{i_k}) }
But the trace of each term essentially produces a factor $\beta^p$,
where $p$ denotes the number of loops formed by the configuration 
of the faces of $R'^*$. Choosing $x$ to be the isotropic point $x=z$,
where $W_i= 1_i + U_i$, we have simply calculated
the partition function of the dense loop model, whose
configurations are the $2^{NM}$ face configurations obtained
by taking either
\eqn\either{\figbox{2.cm}{ig.eps}\ \ \ 
{\rm or}\ \ \ \figbox{2.cm}{ug.eps}} 
on each face of $R'^{*}$, and by attaching a weight $\beta$ per loop.

\newsec{The Bi-colored Dense Loop Model}

In this section, we generalize the dense loop model by using 
the two-color Fuss-Catalan algebra recently introduced by 
Bisch and Jones \FC.

\subsec{The Fuss-Catalan Algebra}

The two-color Fuss Catalan algebra $FC_{2N+2}(\alpha,\beta)$ is defined 
using the domino pictorial representation of previous section, but
its elements must satisfy a further constraint.
The dominos have $2N+2$ top and bottom ends of strings, which 
are painted using two colors, say $a$ and $b$, following
the same pattern $abbaabbaabbaabba..$. The latter ends either with $a$ 
($N$ odd) or with a $b$ ($N$ even). The constraint is that only
ends of the same color can be connected.
In addition to the identity, which satisfies the constraint,
this leads to the following generators:
\eqn\gfuca{ U_i^{(1)} = \figbox{3.5cm}{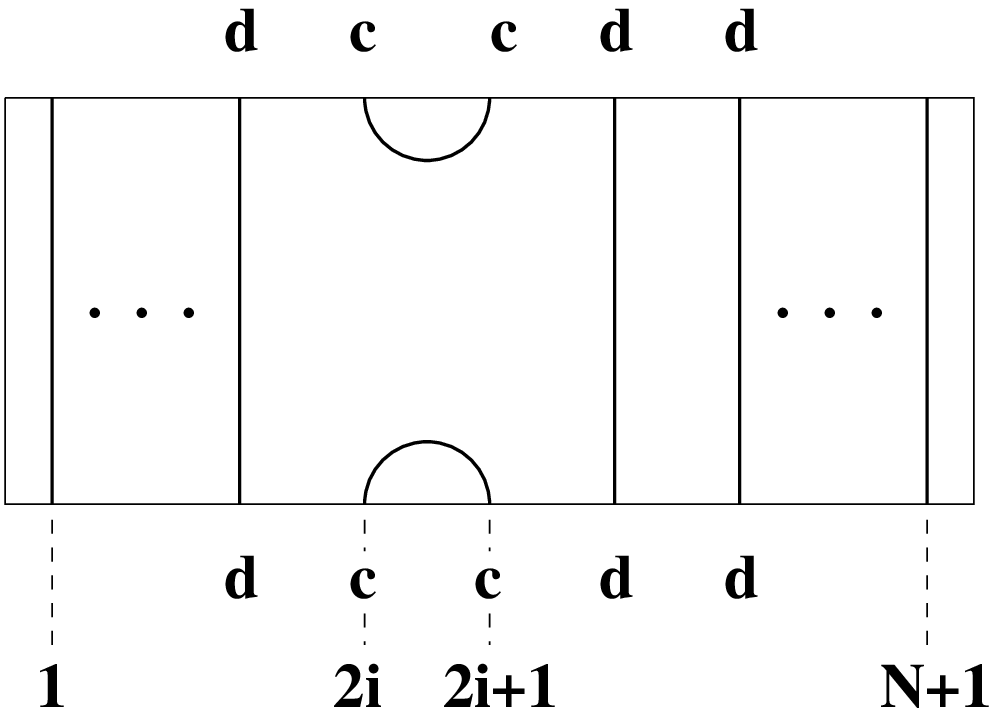}  \qquad U_i^{(2)} = 
\figbox{3.5cm}{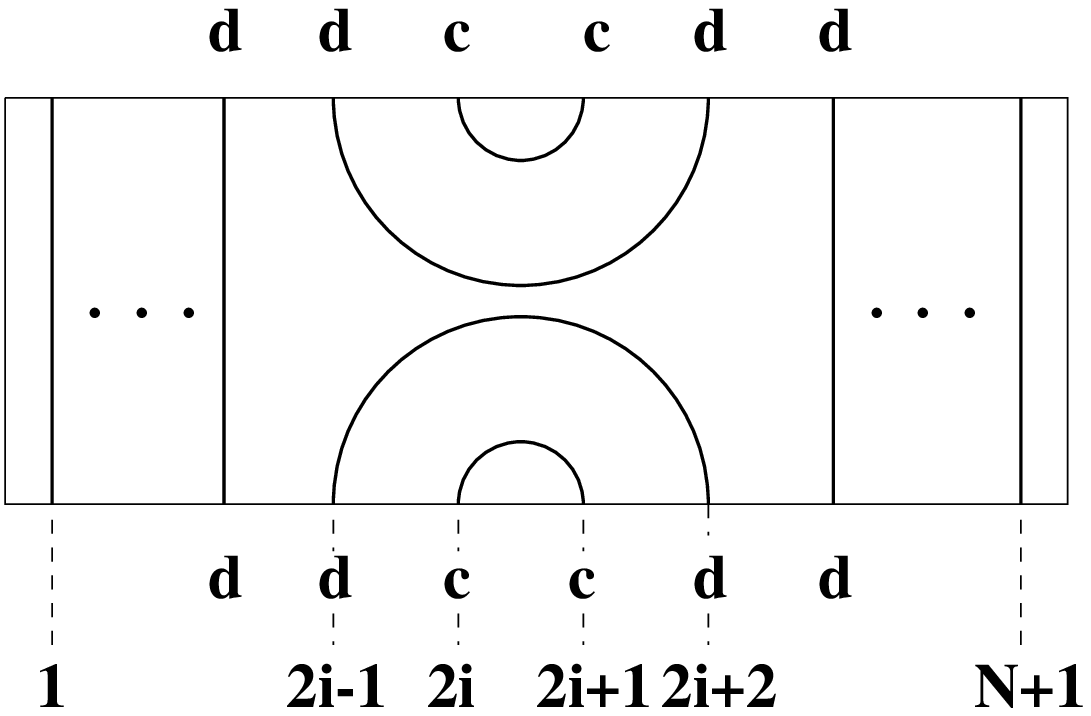}}
for $i=1,2,\ldots,N$.
Note the positions at which these generators act: the position label $i$
of $U_i$ should be thought of as that of the center of the segment 
$2i,2i+1$, joining the corresponding string ends. When $i$ is even, 
the central strings at positions $2i$ and $2i+1$ have color $a$.
It is $b$ when $i$ is odd.

The pictorial representation translates into the following relations,
where loops of color $a$ (resp. $b$) receive a weight $\alpha$
(resp. $\beta$).
\eqn\relfuca{\encadremath{
\eqalign{ \big( U_i^{(1)}\big)^2 &= \left\{ \matrix{ 
\alpha U_i^{(1)} & \ \ {\rm for} \ \ i \ {\rm even} \cr
\beta U_i^{(1)} & \ \ {\rm for} \ \ i \ {\rm odd} \cr}
\right. \cr
\big( U_i^{(2)}\big)^2 &= \alpha \beta U_i^{(2)} \cr
U_{i}^{(1)} U_i^{(2)} &= U_i^{(2)}U_{i}^{(1)} =\left\{ \matrix{ 
\alpha U_i^{(2)} & \ \ {\rm for} \ \ i \ {\rm even} \cr
\beta U_i^{(2)} & \ \ {\rm for} \ \ i \ {\rm odd} \cr}
\right. \cr 
U_{i}^{(1)} U_{i+1}^{(1)} &= U_{i+1}^{(1)} U_{i}^{(1)}\cr
U_{i}^{(1)} U_{i\pm 1}^{(2)} U_{i}^{(1)} &= U_{i}^{(1)} U_{i\pm 1}^{(1)}\cr
U_{i}^{(2)} U_{i\pm 1}^{(1)} U_{i}^{(2)} &= \left\{ \matrix{ 
\alpha U_i^{(2)} & \ \ {\rm for} \ \ i \ {\rm even} \cr
\beta U_i^{(2)} & \ \ {\rm for} \ \ i \ {\rm odd} \cr}
\right. \cr
U_{i}^{(2)} U_{i\pm 1}^{(2)} U_{i}^{(2)} &= U_{i}^{(2)}\cr}}}

Note that $U^{(2)}$ is a generator of $TL_{N+1}(\alpha \beta)$.
Note also that we may write more cubic relations, as consequences of
\relfuca, namely
\eqn\morela{\eqalign{
U_{i}^{(1)} U_{i\pm 1}^{(2)} U_{i}^{(2)} &= U_{i\pm 1}^{(1)} U_{i}^{(2)}\cr
U_{i}^{(2)} U_{i\pm 1}^{(2)} U_{i}^{(1)} &= U_{i}^{(2)} U_{i\pm 1}^{(1)}\cr}}

\subsec{The Yang-Baxter Solution}

Let us look for solutions of the Yang-Baxter equation 
\ybe\-\norma\-\imposi\ in
the form
\eqn\forfuca{W_i(x) = 1_i + a(x) U_i^{(1)} +b(x) U_{i}^{(2)} }
Up to an interchange $i \leftrightarrow i+1$,
we can always assume that $i$ is even in \ybe.
Using the relations \relfuca, we have reexpressed \ybe\ as
a linear combination of reduced elements of $FC_{2N+2}(\alpha,\beta)$.
Their coefficients must vanish, leading to the following equations
(for simplicity, we have set $f(x)=f$, $f(xy)=f'$ and $f(y)=f''$
for $f=a,b$)
\eqn\vanish{\eqalign{
U_i^{(1)} \ &: \ a+a''+\alpha a a'' - a'=0\cr
U_{i+1}^{(1)}&: \ a+a''+\beta a a'' - a'=0\cr
U_i^{(2)} \ &: \ (1+\alpha a'')b+(1+\alpha a)b''+(\alpha \beta
+\alpha a'+b')bb'' -b'=0 \cr
U_{i+1}^{(2)}&: \ (1+\beta a'')b+(1+\beta a)b''+(\alpha \beta
+\beta a'+b')bb'' -b'=0 \cr
U_i^{(1)}U_{i+1}^{(2)}&-U_i^{(2)}U_{i+1}^{(1)}:\cr
&\ \  ab'-a'b-bb'a''-\beta ba'a''=0\cr
U_{i+1}^{(1)}U_{i}^{(2)}&-U_{i+1}^{(2)}U_{i}^{(1)}:\cr
&\ \  a'b''-a''b'+ab'b''+\alpha ba'a''=0\cr}}
where we have indicated the corresponding element of 
$FC_{2N+2}(\alpha,\beta)$ in factor.
Moreover, the normalization condition \norma\ yields, for even $i$
(for simplicity, we have set $f(1/x)={\bar f}$, for $f=a,b$)
\eqn\inverab{\eqalign{
a +{\bar a}+\alpha a {\bar a} &=0\cr
(1+\alpha {\bar a})b+(1+\alpha a){\bar b}+\alpha\beta b {\bar b} &=0}}

As we are looking for non-trivial solutions, we want $a\neq 0$,
hence from the first two lines of \vanish, we must have 
\eqn\egalb{\alpha = \beta}
Expanding at first order in $\epsilon$,
with $y=e^\epsilon/x$, and using $a(1)=0$ from \imposi,
we find the differential equation
\eqn\difa{ xa'(x) =a'(1) (1+\alpha a(x)) }
easily solved as $a(x)=(x^\gamma-1)/\alpha$, where we
can set $\gamma=\alpha a'(1)$ to $1$ without restrictions (by setting
$x=e^{\gamma u}$ as before).
This gives 
\eqn\sola{ a(x) = { x-1 \over \alpha} }
Substituting this into the fifth equation of \vanish, we get
\eqn\dib{ (x-1-(y-1)b(x)) b(xy) -y(xy-1) b(x) =0 } 
Expanding to first order in $\epsilon$, where $y=e^\epsilon$,
we find the differential equation
\eqn\difb{ x(x-1) b'(x) -b(x)^2 -(2x-1) b(x) =0}
which may be recast into $\big(x(x-1)/b(x)\big)'=-1$, hence we have
\eqn\bfor{ b(x) ={x(x-1) \over \mu - x} }
for some integration constant $\mu$. It is easy to see that $\mu$ is 
not determined by \dib.
The only equation left is the third line of \vanish, which yields
\eqn\firli{ y b(x) +x b(y) +b(x)b(y)(\alpha^{2}-1+xy) +b(xy)\big( 
b(x)b(y)-1\big) = 0}
Substituting \bfor\ into this, we finally get
\eqn\deterb{ {xy (xy+1)(\alpha^2-1-\mu)\over (\mu-x)(\mu-y)} = 0}
hence $\mu=\alpha^2-1$, and finally
\eqn\valb{ b(x) = {x(x-1) \over \alpha^2-1 - x} } 

To summarize, we have found a new hyperbolic solution
of the Yang-Baxter equation of the form
\eqn\forfoca{\encadremath{
W_i(x) = 1_i + {x-1 \over \alpha} U_i^{(1)} 
+{x (x-1) \over \alpha^{2}-1 -x} U_i^{(2)}  }}
where $U_{i}^{(1,2)}$ are the generators of the two-color
Fuss-Catalan algebra $FC_{2N+2}(\alpha,\alpha)$, subject to
\relfuca\ with $\alpha=\beta$. Note that we must take 
$\alpha^2\neq 2$, in order for the condition \imposi\ to hold.
Moreover, if we insist on having positive Boltzmann weights,
we must take $\alpha^2>2$ and the range of physical spectral
parameters is given by
\eqn\physpar{ 1<x<\alpha^2-1}

\subsec{The Loop Model}

The solution \forfoca\  yields a new integrable model, which we now
describe in terms of dense loops on the square lattice.
First note that the pictorial representation \gfuca\ may be simplified
by representing only the local part of the generators which is not the 
identity. This suggests to introduce the following face configurations
\eqn\sqconf{ 1_i = \figbox{2.cm}{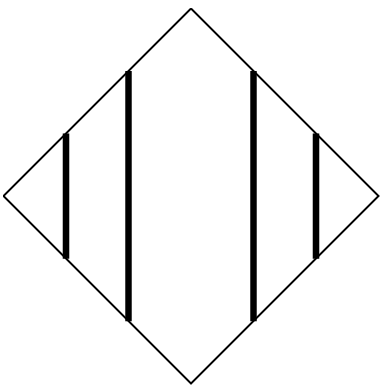} \qquad
U_i^{(1)} = \figbox{2.cm}{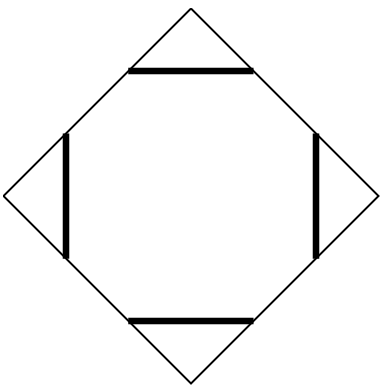} \qquad
U_i^{(2)} = \figbox{2.cm}{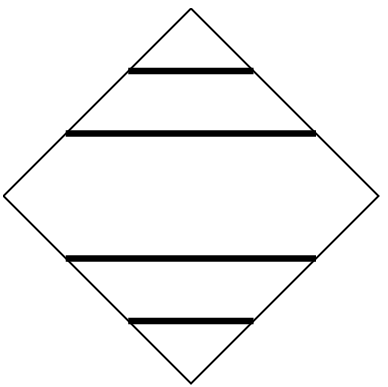}}
where the inner strings have color $a$ if $i$ is even, $b$ if $i$ is 
odd. Note that $\alpha=\beta$, hence we need not represent the 
strings with different colors, as eventually all the loops will 
receive the same weight $\alpha$.

The partition function of the model on a rectangle with $M\times N$ 
faces is obtained by picking any of the 
three configurations of \sqconf\ on each face of the rectangle,
and summing over all $3^{NM}$ possible configurations of the 
rectangle these generate. Each such configuration is weighed by a
factor $a(x)^k b(x)^m$, where $k$ and $m$ denote the 
total numbers of faces of respectively the second and third types of \sqconf.
Imposing periodic boundary conditions on the top and bottom of the
rectangle, the partition function is then obtained by taking the trace 
of the corresponding products of $U_i$'s. 
This Markov trace is defined on any reduced element
as $\alpha^p\beta^q$, where $p$ and $q$ denote respectively the total numbers 
of loops of color $a$ and $b$, obtained by identifying the top and bottom 
of each string. 
The definition is extended to any element by linearity.
As we have set $\alpha=\beta$, each loop configuration receives an 
extra weight $\alpha^p$, where $p$ is the total number of loops formed.
To summarize, the partition function reads 
\eqn\pfuca{ Z = \sum_{{\rm face}\ {\rm configurations}\atop
\figbox{.4cm}{iig.eps}, \figbox{.4cm}{iug.eps}\ {\rm or}\figbox{.4cm}{uug.eps}}
a(x)^k b(x)^m \alpha^p}

\fig{A configuration of the two-color loop model. Each face is in
one of the three face configurations of \sqconf.}{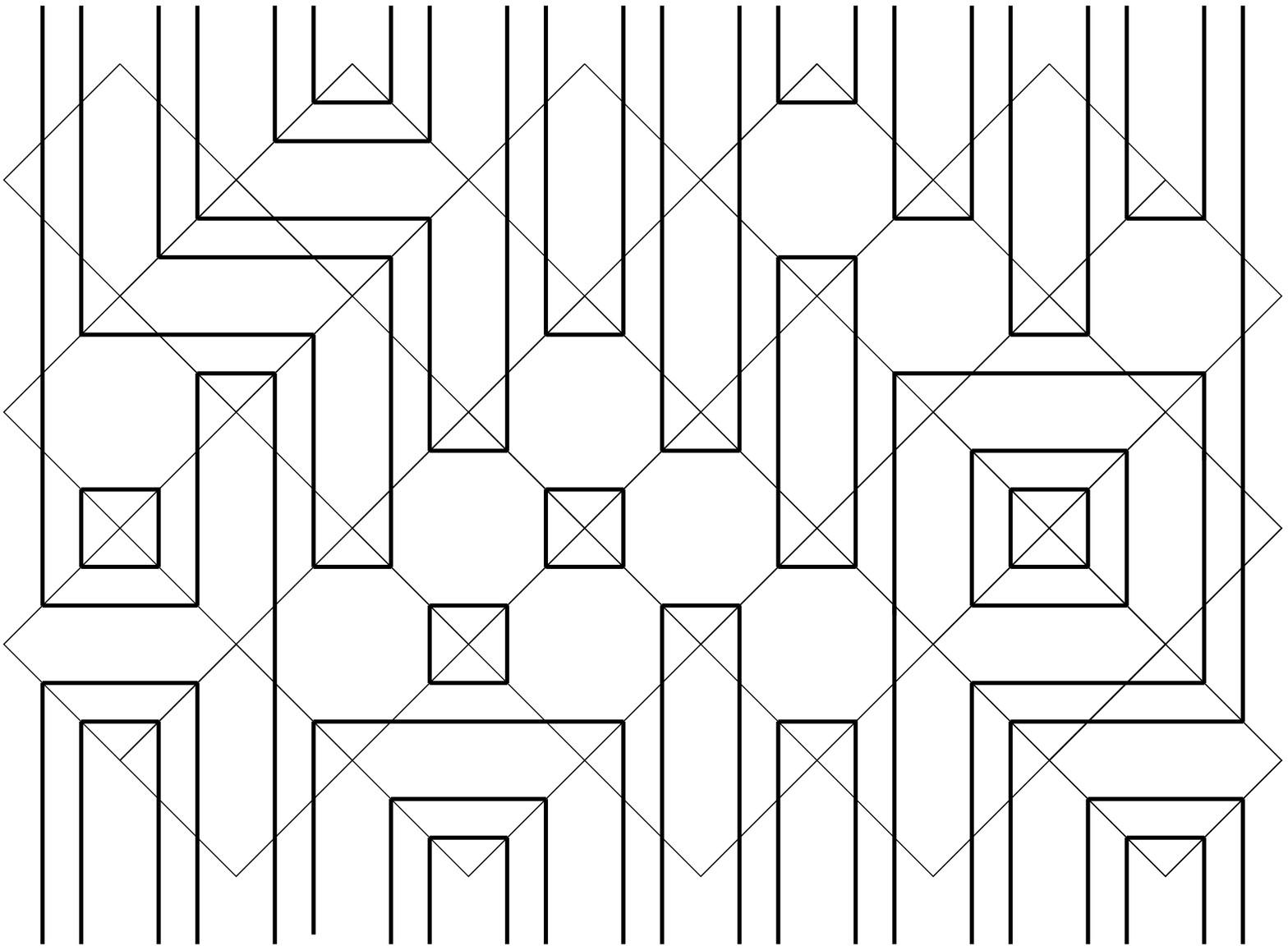}{7.cm}
\figlabel\confuca

We have represented a typical configuration in Fig.\confuca.
Taking for $x$ the isotropic value $x_*=\sqrt{\alpha^{2}-1}$, we have
$b(x_*)=1$ and $a(x_*)=(\sqrt{\alpha^{2}-1} -1)/\alpha$. At this 
point, we simply have a model of loops, with partition function
\eqn\sipart{Z = \sum_{{\rm face}\ {\rm configurations}\atop
\figbox{.4cm}{iig.eps}, \figbox{.4cm}{iug.eps}\ {\rm or}\figbox{.4cm}{uug.eps}}
\big(\sqrt{\alpha^2-1}-1\big)^k \alpha^{p-k}}
where $k$ is the total number of faces of the second type in \sqconf\ 
and $p$ the total number of loops.

Note that in this model the loops are fully packed and are
colored with two colors,
according to the pattern of the Fuss-Catalan algebra.

Note also the existence of a crossing symmetry for the Boltzmann 
weights \forfoca, expressing the global covariance of the
model under a rotation of $90^\circ$:
\eqn\crossi{ {\bar W}_i(x_*^2/x) ={x_*^2-x \over x(x-1)}\  W_i(x) }
where $x_*=\sqrt{\alpha^2-1}$, and
the bar stands for the effect of a rotation of $90^\circ$,
namely ${\bar 1_i}=U_i^{(2)}$, ${\bar U_i}^{(2)}=1_i$, and 
${\bar U_i}^{(1)}=U_i^{(1)}$. In particular, we have
${\bar W_i}(x_*)=W_i(x_*)$, hence the model \sipart\ is invariant under 
rotation of $90^\circ$. 

\newsec{The Multi-colored Dense Loop Model}

This section is an extension of the results of Sect.3 to the case 
of multi-colored loops. We find new hyperbolic solutions of the 
Yang-Baxter equation, based on the $k$-color Fuss-Catalan algebra.

\subsec{Higher Fuss-Catalan Algebras}
\par
Higher Fuss-Catalan algebras have been introduced in \FC, within the 
same type of pictorial representation as before, but now corresponding
to painting the strings with $k$ distinct colors $a_1,a_2,\ldots,a_k$, 
and attaching a weight $\alpha_j$ per loop of color $a_j$. These are 
the $k$-color Fuss-Catalan algebras, denoted by 
$FC_{k(N+1)}(\alpha_1,\alpha_2,\ldots,\alpha_k)$.
Each domino has now $k(N+1)$ strings whose ends are painted following
the pattern: $a_1a_2\ldots a_k a_k a_{k-1}\ldots a_2 a_1 a_1 a_2\ldots$,
both on the top and bottom ends.
If $N$ is odd the pattern ends with $a_1$, if $N$ is even, with $a_k$.
The constraint is now that a string may only connect points of the
same color.

In analogy with Sect.3, we have found the $k$ following types of generators
for $m=1,2,\ldots,k$
\eqn\ktyp{ U_i^{(m)} = \figbox{5.cm}{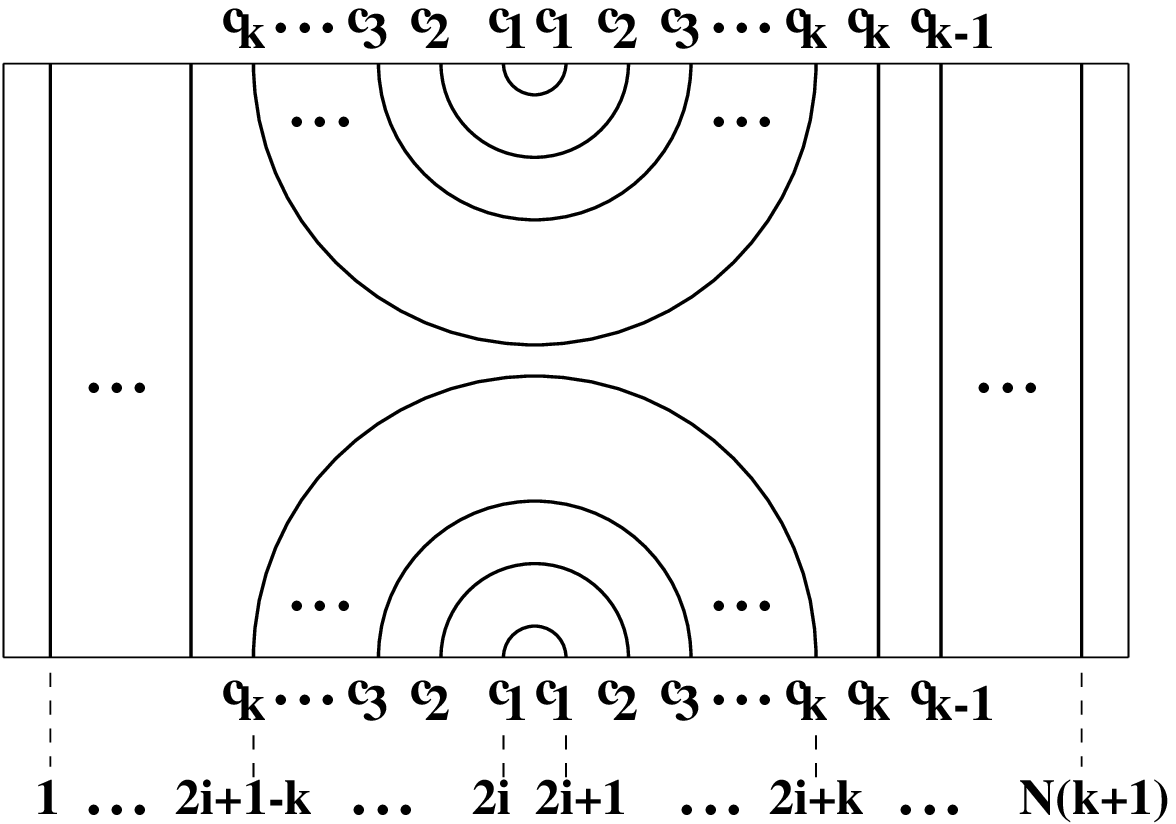} } 
where $c_j=a_j$ if $i$ is even, and $c_j=a_{k+1-j}$ if $i$ is odd.
These generators are all un-normalized projectors, and they satisfy 
the quadratic relations for $1\leq m\leq p\leq k$
\eqn\quafu{ U_i^{(m)} U_i^{(p)}= U_i^{(p)} U_i^{(m)}
=\rho_i{(m)}  U_i^{(p)}}
where 
\eqn\roip{\rho_i(p) = \left\{ \matrix{ 
\alpha_1\alpha_2\ldots\alpha_{p} & \ \ {\rm if}\ i \ {\rm even}\cr
\alpha_k\alpha_{k-1}\ldots\alpha_{k+1-p} & \ \ {\rm if}\ i \ {\rm 
odd}\cr}\right. }

The generators $U_i^{(m)}$ are local, hence 
\eqn\commop{  U_i^{(m)} U_j^{(p)}= U_j^{(p)} U_i^{(m)} }
whenever $\vert i-j\vert >1$, or when $\vert i-j\vert=1$, and 
$m+p\leq k$. 

\fig{The relation $U_i^{(m)} U_{i+1}^{(p)} U_i^{(q)} 
= \rho_i(k-p) U_i^{(m)} U_{i+1}^{(k-q)}$ for $m\geq q$. 
The generators
are represented by squares of respective size $m,p,q$.
The $m$ and $q$ boxes are first crushed into each other,
by eliminating the $k-p$ loops created between them,
and replacing them with the weight $\rho_i(k-p)$.}{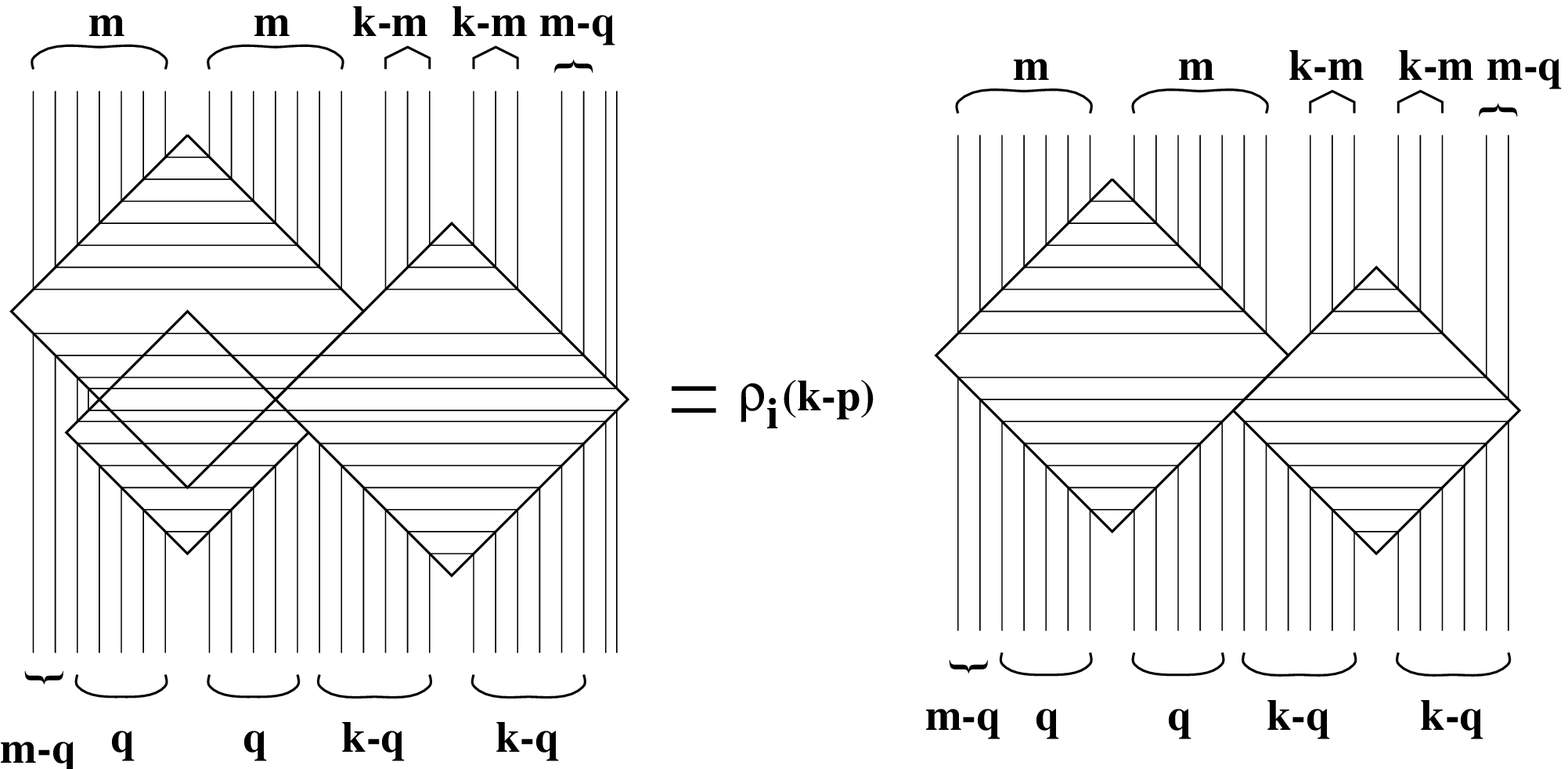}{9.cm}
\figlabel\projo

When these operators do not commute, it is easy to show that they 
always satisfy cubic relations generalizing \tla, \relfuca\ and \morela.
These read
\eqn\cubicu{ U_i^{(m)} U_{i\pm 1}^{(p)} U_i^{(q)} = \rho_i(k-p) \left\{
\matrix{ U_i^{(m)} U_{i\pm 1}^{(k-q)} & \ \ {\rm if} \ m\geq q \cr
U_{i\pm 1}^{(k-m)} U_i^{(q)} & \ \ {\rm if} \ m\leq q \cr}\right.}
with $\rho_i(p)$ as in \roip. The first relation is explained in 
Fig.\projo.
Note that the relations \quafu\ and \cubicu\ are not independent. 
For instance, assume that $m \geq q$ in \cubicu. Then multiplying the 
cubic relation by $U_i^{(r)}$ from the left, for $r>q$, and using the
quadratic relations \quafu, we obtain the equation \cubicu\ with
$r$ substituted for $m$. This shows that we only need to write the
relation for the smallest possible $m \geq q$, namely $m=q$, and
that all the others will be consequences.

The $k$-color Fuss-Catalan algebra is therefore defined by the 
generators 
$1\equiv U_i^{(0)}$ and  $U_i^{(m)}$, $m=1,2,\ldots,k$ and 
$i=1,2,\ldots,N$, subject to the relations
\eqn\fucalge{\encadremath{
\eqalign{U_i^{(m)} U_i^{(p)}&= U_i^{(p)} 
U_i^{(m)}=\rho_i(m) U_i^{(p)}\cr
U_i^{(m)} U_j^{(p)}&= U_j^{(p)} U_i^{(m)}\ \ {\rm if}\ \vert 
i-j\vert>1\ {\rm or}\  j=i\pm 1\ {\rm and}\ m+p\leq k\cr
U_i^{(m)} U_{i\pm 1}^{(p)} U_i^{(m)} &= \rho_i(k-p) U_i^{(m)} 
U_{i\pm 1}^{(k-m)}\ \ {\rm for}\ m+p>k \cr}}}

\subsec{The Yang-Baxter Equation: I-Equations}

Let us now look for a solution to the Yang-Baxter equation 
\ybe\-\norma\-\imposi\ in the form
\eqn\forwfc{ W_i(x)= 1_i +\sum_{m=1}^k a_m(x) U_i^{(m)} }
where $a_m(x)$ are some functions of $x$ to be determined, and
the $U_i^{(m)}$ are the generators of 
$FC_{k(N+1)}(\alpha_1,\ldots,\alpha_k)$.

As before, we simply have to expand \ybe\ onto products of $U$'s,
which we must rearrange using the relations \quafu\ and \cubicu.
Like in the case $k=2$,
let us restrict ourselves to even $i$, and denote by 
$\rho_m=\rho_i(m)=\alpha_1\alpha_2\ldots\alpha_m$,
and by $\sigma_m=\rho_{i+1}(m)=\alpha_k\alpha_{k-1}\ldots\alpha_{k+1-m}$, with 
$\rho_0=\sigma_0=1$.
We list below the monomials in the $U$'s and the corresponding 
vanishing coefficients (as before, we write $f,f',f''$ for 
$f(x),f(xy),f(y)$ respectively). We start with the first degree 
monomials:
\eqn\firdeg{\eqalign{ 
U_i^{(m)},\ 1\leq m<k&:\ (\sum_{p=0}^{m-1} \rho_p a_p'') a_m 
+(\sum_{p=0}^{m-1} \rho_p a_p) a_m''+\rho_m a_m a_m''-a_m' =0 \cr
U_{i+1}^{(m)},\ 1\leq m<k&:\ (\sum_{p=0}^{m-1} \sigma_p a_p'') a_m 
+(\sum_{p=0}^{m-1} \sigma_p a_p) a_m''+\sigma_m a_m a_m''-a_m' =0 \cr
U_i^{(k)}\ \ \ \ \ &:\ (\sum_{p=0}^{k-1} \rho_p a_p'') a_k 
+(\sum_{p=0}^{k-1} \rho_p a_p) a_k''+(\sum_{p=0}^{k-1}\rho_p a_{k-p}') 
a_k a_k''-a_k' =0 \cr
U_{i+1}^{(k)}\ \ \ \ &:\ (\sum_{p=0}^{k-1} \sigma_p a_p'') a_k 
+(\sum_{p=0}^{k-1} \sigma_p a_p) a_k''+(\sum_{p=0}^{k-1}\sigma_p a_{k-p}')
a_k a_k''-a_k' =0 \cr}}
where the extra terms for $m=k$ arise from the cubic relations
$U_i^{(k)}U_{i+1}^{(k-p)} U_i^{(k)}=\rho_p U_i^{(k)}$, and
$U_{i+1}^{(k)}U_i^{(k-p)}U_{i+1}^{(k)}=\sigma_{p}U_{i+1}^{(k)}$, for 
$p=1,2,\ldots,k$.
The quadratic terms $U_i^{(m)}U_{i+1}^{(l)}$ or $U_{i+1}^{(m)}U_i^{(l)}$
are distinguished according to whether they commute ($m+l\leq k$) or
not ($m+l>k$). Actually, the maximally commuting case $m+l=k$ must
be treated separately.
The commuting terms read:
\eqn\secdeg{\eqalign{U_i^{(m)}U_{i+1}^{(l)}&:\ 
(\sum_{p=0}^{m-1} \rho_p a_p'') 
a_l'a_m''+(\sum_{p=0}^{m-1} \rho_p a_p) a_m'a_l''\cr
&-(\sum_{p=0}^{l-1} \sigma_p a_p'') a_la_m'
-(\sum_{p=0}^{l-1} \sigma_p a_p) a_m'a_l''=0\cr}}
for $m+l< k$.
When $m+l=k$, we get some extra terms:
\eqn\extracom{\eqalign{&(\sum_{p=0}^{m-1} \rho_p a_p)a_{k-m}'a_m''+
(\sum_{p=0}^{m-1} \rho_p a_p'')a_{k-m}'a_m+
(\sum_{p=0}^m \rho_p a_{k-p}')a_ma_m''\cr
&-(\sum_{p=0}^{k-m-1} \sigma_p a_p) a_{k-m}''a_m'
-(\sum_{p=0}^{k-m-1} \sigma_p a_p'') a_m'a_{k-m}
-(\sum_{p=0}^{k-m} \sigma_p a_{k-p}')a_{k-m}a_{k-m}''=0\cr}} 

The non-commuting ones are 
\eqn\noncom{\eqalign{U_i^{(m)}U_{i+1}^{(l)}&:\ 
(\sum_{p=0}^{k-l} \rho_p a_p'')a_ma_l'-(\sum_{p=0}^{k-m} \sigma_p 
a_p'')a_la_m'\cr
&+(\sum_{p=l+1}^{k} \rho_{k-p} a_p')a_ma_{k-l}''-
(\sum_{p=m+1}^{k} \sigma_{k-p} a_p')a_la_{k-m}''=0\cr
U_{i+1}^{(m)}U_i^{(l)}&:\ 
(\sum_{p=0}^{k-l} \sigma_p a_p)a_m''a_l'-(\sum_{p=0}^{k-m} \rho_p 
a_p)a_l''a_m'\cr
&+(\sum_{p=l+1}^{k} \sigma_{k-p} a_p')a_m''a_{k-l}-
(\sum_{p=m+1}^{k} \rho_{k-p} a_p')a_l''a_{k-m}=0\cr}}
for $m+l>k$.

\subsec{The Yang-Baxter Equation: II-Solutions}
 
The first two equations of \firdeg\ have non-trivial solutions with $a_m\neq 0$ only if 
both are the same, namely $\sigma_p = \rho_p$ for $p=1,2,\ldots,k$,
which implies that
\eqn\symea{ \alpha_m = \alpha_{k+1-m} \ \ \ 
{\rm for} \ \ 1\leq m\leq k}
This implies also that the two equations of \noncom\ are equivalent.
Note that the first equation of \firdeg\ for $m=1$ coincides with the first line of 
\vanish, with $\alpha=\beta$ replaced by $\alpha_1=\alpha_k$. Hence
we have the solution
\eqn\solaone{ a_1(x) = {x^{r_1}-1 \over \alpha_1} }
where $r_1$ can be set to $1$ as usual (we will however keep
$r_1$ in the formulas, for the sake of uniformity).
For $m=2$, we have
\eqn\mto{(1+\alpha_1 a_1(y))a_2(x)+(1+\alpha_1 
a_1(x))a_2(y)+\alpha_1\alpha_2a_2(x) a_2(y) -a_2(xy)=0}
Writing $a_2(x)=x^{r_1} m_2(x)$, and using the value \solaone, we find
\eqn\mars{ m_2(x)+m_2(y)+\alpha_1\alpha_2m_2(x)m_2(y)-m_2(xy)=0}
This is again the first equation of \vanish, with the substitution 
$\alpha\to \alpha_1\alpha_2$.
We deduce the solution $m_2(x)=(x^{r_2}-1)/(\alpha_1\alpha_2)$, where 
$r_2$ is an arbitrary nonzero number, hence
\eqn\qato{a_2(x)={1\over \rho_2}x^{r_1}(x^{r_2}-1)} 
Proceeding by induction, 
assume $a_p=x^{r_1+r_2+\ldots+r_{p-1}}(x^{r_p}-1)/\rho_p$, for all 
$p\leq q-1$, and some non-vanishing numbers $r_1=1,r_2,\ldots,r_{q-1}$. 
Then, setting $a_q=x^{r_1+r_2+\ldots+r_{q-1}}m_q(x)$,
and noting that 
\eqn\sumimp{\sum_{p=0}^{q-1} \rho_p 
a_p(x) = x^{r_1+r_2+\ldots+r_{q-1}} }
the equation for $m_q$ is nothing but
again the first equation of \vanish, with $\alpha\to \rho_q$.
We therefore get $m_q=(x^{r_q}-1)/\rho_q$ for some 
non-vanishing number $r_q$, and 
\eqn\valala{ a_m(x) = {1\over \rho_m}
x^{r_1+r_2+\ldots+r_{m-1}}(x^{r_m}-1) }
for all $m=1,2,\ldots,k-1$, and $r_i\neq 0$, $i=1,2,\ldots,k-1$.

Let us now turn to the commuting second order term 
equations \secdeg. The solutions \valala\
satisfy them automatically. Indeed, using \symea\ and \sumimp,
\secdeg\ read
\eqn\redthy{ \eqalign{&x^{r_1+r_2+\ldots+r_{m-1}}a_l'a_m''
+y^{r_1+r_2+\ldots+r_{m-1}}a_ma_l'+\rho_ma_m a_l'a_m''\cr
&-x^{r_1+r_2+\ldots+r_{l-1}}a_m'a_l''-y^{r_1+r_2+\ldots+r_{l-1}}
a_la_m'-\rho_l a_la_m'a_l''\cr
&= {1\over \rho_m\rho_l} (xy)^{r_1+\ldots+r_{m-1}+r_1+\ldots+r_{l-1}}
\bigg( ((xy)^{r_l}-1)(x^{r_m}-1+y^{r_m}-1)\cr
&-((xy)^{r_m}-1)(x^{r_l}-1+y^{r_l}-1)
+((xy)^{r_l}-1)(x^{r_m}-1)(y^{r_m}-1)\cr
&\ \ \ \ \ \ \ \ \ \ \ \ \ \ 
-((xy)^{r_m}-1)(x^{r_l}-1)(y^{r_l}-1)\bigg)=0\cr}}
where we have substituted the values \valala\ (note that $m+l\leq k$
and $m,l\geq 1$, hence $m,l\leq k-1$ and \valala\ applies). 

Let us now use the highest non-trivial of the non-commuting
equations \noncom, for $m=k$ and $l=k-1$ (for $m=l=k$ the coefficient
is identically zero). This gives 
\eqn\isgiv{(a_{k-1}(x)-a_k(x)a_1(y)) a_k(xy) = y a_k(x)a_{k-1}(xy) }
Introducing the ratio
\eqn\ratiphi{ \varphi_1(x) = x {a_{k-1}(x)\over a_k(x)} }
this becomes
\eqn\beco{ \varphi_1(xy)-\varphi_1(x) = -{x(y-1)\over \rho_1}}
Setting $y=e^\epsilon$ and expanding to first order in $\epsilon$,
we get $\varphi_1'(x) = -1/\rho_1$, easily integrated into
\eqn\valuphi{ \varphi_1(x) = {\mu - x \over \rho_1} }
for some integration constant $\mu$.
{}From the definition \ratiphi, we deduce 
\eqn\valak{ a_k(x) ={\rho_1\over \rho_{k-1}}
x^{r_1+r_2+\ldots+r_{k-2}+1 }{x^{r_{k-1}}-1\over \mu-x}  }
Note that $\mu\neq 1$ in order for the initial condition $a_k(1)=0$ to 
be satisfied. It is easy to check that \valuphi\ satisfies the initial 
equation \beco\ trivially, for any $\mu$.

We can now substitute the forms \valala\-\valak\ into the remaining 
equations, to further fix the parameters. Let us now  
use \noncom\ for $m=k$ and $l=k-2$.
The equation reads
\eqn\twosub{ {a_{k-2}(x)\over a_k(x)} =a_2(y) (1+\rho_1 
{ a_{k-1}(xy)\over a_k(xy)}+(1+\rho_1 a_1(y)+\rho_{2}a_2(y)) 
{a_{k-2}(xy)\over a_k(xy)} }
Using the value \ratiphi, the identity $1+\rho_1 
a_1(y)+\rho_{2}a_2(y)=y^{r_1+r_2}$ from \sumimp, and setting 
$\varphi_2(x)=x^{r_1+r_2}a_{k-2}(x)/ a_k(x)$, we get
\eqn\phicom{ \varphi_2(xy)-\varphi_2(x) = -{\mu\over \rho_2}
x^{r_2}(y^{r_2}-1) }
Setting $y=e^\epsilon$ and expanding to first order in $\epsilon$,
we find the differential equation
\eqn\diphi{ \varphi_2'(x) = -{\mu\over \rho_2}r_2 x^{r_2-1} }
easily integrated as
\eqn\diphi{ \varphi_2(x) = {\mu\over \rho_2}(\mu^{r_2}-x^{r_2}) }
as $\varphi_2(x)$ vanishes at $x=\mu$. 
We may now proceed by induction. Assume 
\eqn\hypophi{\varphi_p(x) = {1\over \rho_p}\mu^{r_1+r_2+\ldots+r_{p-1}}
(\mu^{r_p}-x^{r_p}) }
for all $p\leq q-1$. Then writing the equation \noncom\ for
$m=k$ and $l=k-q$ and dividing it by $a_k a_k'$, we get
\eqn\kmqd{\eqalign{
\varphi_q(xy)-\varphi_q(x) &= -x^{r_1+\ldots+r_q}a_q''
\big(1+{\rho_1\over xy}\varphi_1(xy)+\ldots+{\rho_{q-1}\over 
(xy)^{r_1+\ldots+r_{q-1}}} \varphi_{q-1}(xy) \big) \cr
&= -{1\over \rho_q}\mu^{r_1+\ldots+r_q}x^{r_p}(y^{r_p}-1)\cr}}
where we have used the induction hypothesis \hypophi.
Setting $y=e^\epsilon$, and expanding to first order in $\epsilon$,
we get a differential equation, easily integrated as
\eqn\intephi{ \varphi_q = 
{1\over \rho_q}\mu^{r_1+r_2+\ldots+r_{q-1}}
(\mu^{r_q}-x^{r_q}) }
which therefore holds for all $q=1,2,\ldots,k-1$.
Note the following remarkable property of this solution,
namely that
\eqn\remapro{ 1+ {\rho_1\over x^{r_1}}\varphi_1(x) +
{\rho_2\over x^{r_1+r_2}}\varphi_2(x) +\ldots
+{\rho_m\over x^{r_1+\ldots+r_m}}\varphi_m(x)=
\big({\mu\over x}\big)^{r_1+\ldots+r_m} }
This will be used extensively later.

We have now two values for the ratios \ratiphi. One is \intephi,
the other is obtained by substituting \valala\ into the
definition \intephi. Both agree if
\eqn\agree{{1\over \rho_q}\mu^{r_1+r_2+\ldots+r_{q-1}}
(\mu^{r_q}-x^{r_q}) = {\rho_{k-1}\over\rho_1\rho_{k-q}}
{x^{r_1+\ldots+r_q} (x^{r_{k-q}}-1)(\mu-x)\over
x^{r_{k-q}+\ldots+r_{k-2}} (x^{r_{k-1}}-1)} }
for all $q=1,2,\ldots,k-1$.
The simplest way to analyze these equations is to take 
the ratio of two consecutive ones, leading to
\eqn\consec{ x^{r_{k-q}}(x^{r_q}-\mu^{r_q}) (x^{r_{k-q-1}}-1)=
{\alpha_q^{2}\over \mu^{r_{q-1}}} 
x^{r_{q}}(x^{r_{q-1}}-\mu^{r_{q-1}}) (x^{r_{k-q}}-1) }
for $q=2,3,\ldots,k-1$.
The $r$'s are all non-zero real numbers. By inspection of
the 16 possibilities for the signs of 
$(r_{q-1},r_q,r_{k-q-1},r_{k-q})$, we see that only two do
not lead to contradictions, namely
\item{(i)} $r_{q-1}=r_q=r_{k-q}=r_{k+1-q}$, and $\mu=\alpha_q^{2}$.
\item{(ii)} $r_{q-1}=-r_q$, $r_{k-q}=-r_{k+1-q}$, and $\alpha_q^{2}=1$.
\par
\noindent{}Note that in both cases we have 
\eqn\resalp{ \alpha_q^{2}=\mu^{(r_q+r_{q-1})/2}}
We will discuss these solutions later, let us first write all
remaining equations.

Let us turn to  \extracom. Dividing it by
$a_k'$, and using \intephi, it reads
\eqn\comextra{\eqalign{ 
&\big(1+{\rho_1\over xy}
\varphi_1(xy)+\ldots+
{\rho_{m-1}\over (xy)^{r_1+\ldots+r_{m-1}}} 
\varphi_{m-1}(xy) \big) a_ma_m''\cr
&-\big(1+{\rho_1\over xy}\varphi_1(xy)+\ldots+
{\rho_{k-m-1}\over (xy)^{r_1+\ldots+r_{k-m-1}}} 
\varphi_{k-m-1}(xy) \big) a_{k-m}a_{k-m}''\cr
&={1\over \rho_m^2} \mu^{r_1+\ldots+r_{m-1}}
(x^{r_{m}}-1)(y^{r_{m}}-1)\cr
&-{1\over \rho_{k-m}^2} \mu^{r_1+\ldots+r_{k-m-1}}
(x^{r_{k-m}}-1)(y^{r_{k-m}}-1) = 0\cr }}
where we have used the identity \remapro.
This must hold for all $x,y$, hence
\eqn\req{ r_m = r_{k-m} }
for $m=1,2,\ldots,k-1$. In particular, we learn that $r_{k-1}=r_1=1$.
The coefficients must also agree, hence
\eqn\agreco{ \left({\alpha_1\ldots\alpha_m\over
\alpha_1\ldots\alpha_{k-m}}\right)^{2}=
{\mu^{r_1+\ldots+r_{m}}\over\mu^{r_1+\ldots+r_{k-m}}} }
This identity is a direct consequence of \resalp.

To proceed, let us check the other non-commuting second
degree terms \noncom, namely with $m+l>k$, $m,l<k$. 
Let us take $m=k-p$ and $l=k-s$ in \noncom, mutliply it by
$x^{r_1+\ldots+r_{p}+r_1+\ldots+r_{s}}$, and use \remapro\ to get
\eqn\nocoth{\eqalign{&\varphi_s(x)\varphi_p(xy)
-\varphi_s(x)\varphi_p(xy) = \cr
&\big({\mu x\over xy}\big)^{r_1+\ldots+r_{s-1}}x^{r_s}\varphi_p(x) 
a_s(xy) -\big({\mu x\over xy}\big)^{r_1+\ldots+r_{p-1}}x^{r_p}\varphi_s(x) 
a_p(xy)\cr}}
This is easily checked, by substituting the solution \intephi, and using \agreco. 

Finally, let us write the first degree equation \firdeg\ for $m=k$,
using both \sumimp\ and \remapro:
\eqn\lastleast{\eqalign{a_k(x)a_k(xy)a_k(y) 
&\big({\mu\over xy}\big)^{r_1+\ldots+r_{k-1}}+\rho_k a_k(x)a_k(y)
+y^{r_1+\ldots+r_{k-1}} a_k(x)\cr
&-x^{r_1+\ldots+r_{k-1}}a_k(y)-a_k(xy)=0 \cr}}
Substituting the value \valak, we get after some algebra
\eqn\algeplu{\eqalign{
&{(xy)^{1+r_1+\ldots+r_{k-2}}\rho_1(x-1)
(xy-1)(y-1) \over
\rho_{k-1}(\mu-x)(\mu-xy)(\mu-y)}\times \cr
&\times\bigg( 
(xy-1)\big(\mu(x-1)
(y-1)-(\mu-x)(\mu-y)\big)\cr
&+xy(\mu-xy)\big(\alpha_1^{2}(x-1)(y-1)+
(x-1)(\mu-y)+(y-1)(\mu-x)\big)\bigg)\cr}}
where we have used \agreco\ for $m=1$, 
$\rho_1\rho_k/\rho_{k-1}=\alpha_1^{2}$, and the values 
$r_{k-1}=r_1=1$. The last factor can be rearranged into
\eqn\lasfac{\eqalign{(1-\mu)(xy-1)&(\mu-xy)+(\mu-xy)\big( 
(\mu-1)(xy-1)+(1+\mu-\alpha_1^{2})(x+y)\big)\cr
&=(\mu-xy)(1+\mu-\alpha_1^{2})(x+y)\cr}}
This vanishes for all $x,y$ if 
\eqn\latre{ \mu = \alpha_1^{2}-1 }

To summarize, we have found all solutions $a_1(x),a_2(x),\ldots,a_k(x)$ to
all the equations implied by \ybe\-\norma\-\imposi. These solutions read 
\valala\-\valak, with integers $r_i\in \{\pm 1\}$, loop weights
$\alpha_1,\alpha_2,\ldots,\alpha_{k}$, and a parameter 
$\mu$ such that
\eqn\integs{\eqalign{
r_1&=r_{k-1}=1 \cr
{\rm and} &\left\{ \matrix{{\rm either}\ & r_{q-1}=r_q=r_{k-q}=r_{k+1-q}\cr
{\rm or}\ & r_{q-1}=-r_q=-r_{k-q}=r_{k+1-q}\cr} \right.\cr
\mu^{r_{q-1}+r_q\over 2} &= \alpha_q^{2}\cr
\mu &= \alpha_1^{2}-1\cr
\alpha_q &=\alpha_{k+1-q}\cr}}

Let us start from the fundamental solution
$r_1=r_2=\ldots=r_{k-1}=1$, which has 
$\mu=\alpha_1^{2}-1=\alpha_2^{2}=\alpha_3^{2}=\ldots=\alpha_{k-1}^{2}$ 
i.e. 
\eqn\lopwei{\alpha_2=\alpha_3=\ldots=\alpha_{k-1}=\sqrt{\alpha_1^{2}-1}} 
if all $\alpha$'s are positive, and
\eqn\funda{\encadremath{\eqalign{ a_q(x)&= 
{x^{q-1}\over\alpha_1(\alpha_1^{2}-1)^{q-1\over 2}} (x-1) 
\ \ q=1,2,\ldots,k-1\cr
a_k(x) &= {x^{k-1}\over(\alpha_1^{2}-1)^{{k\over 2}-1}} {x-1\over 
\alpha_1^2-1-x}\cr}}}
This solution generalizes \forfoca\ to the case of $k$-color 
Fuss-Catalan algebras. Note that we have positive Boltzmann weights
only if $\alpha^2>2$, and
\eqn\positgen{ 1<x<\alpha^2-1} 

The other solutions can be obtained by elementary ``$q$-excitations"
in which we reverse all values of $r_i$ for $i=q, q+1,\ldots,k-q$.
Indeed, these excitations must be symmetric, as they impose
that some $\alpha_q=\alpha_{k+1-q}$ change its value from
$\mu$ to $1$ or vice versa.
We can apply these excitations at any $q=2,3,\ldots,\ell$, where 
$\ell=[(k+1)/2]$, hence a total of $2^{\ell-1}$ solutions, of the form
$r_1=\ldots=r_{i_1-1}=1$, $r_{i_1}=\ldots=r_{i_2-1}=-1$, ... ,
$r_{i_{s}}=\ldots=r_{\ell}=(-1)^s$, corresponding to $s$ such 
excitations, $s=0,1,\ldots,\ell-1$. 
The simplest such solution occurs for $k=4$, where we have a total of
two solutions
\eqn\solfor{\eqalign{ (1)&\ {\rm fundamental:}\ \  r_1=r_2=r_3=1\qquad
\alpha_1=\alpha_4=\alpha,\ \ \alpha_2=\alpha_3=\alpha^2-1 \cr
&W_i^{(1)}(x)= 1_i + {x-1\over \alpha} U_i^{(1)} + 
{x(x-1)\over \alpha\sqrt{\alpha^2-1} } U_i^{(2)}\cr
&\ \ \ \ \ \ \ \ \ \ \ \ \ \ \ \ 
+{x^2(x-1)\over \alpha(\alpha^2-1) } U_i^{(3)}
+{x^3(x-1)\over (\alpha^2-1)(\alpha^2-1-x) } U_i^{(4)}\cr
(2)&\ {\rm excited:} \ \ r_1=r_3=1,\ r_2=-1 \qquad 
\alpha_1=\alpha_4=\alpha, \ \ \alpha_2=\alpha_3=1 \cr
&W_i^{(2)}(x)= 1_i + {x-1\over \alpha} U_i^{(1)}-
{(x-1)\over \alpha} U_i^{(2)}\cr
&\ \ \ \ \ \ \ \ \ \ \ \ \ \ \ \ 
+{(x-1)\over \alpha} U_i^{(3)}
+{x(x-1)\over (\alpha^2-1-x) } U_i^{(4)}\cr}}
Note that at the self-dual point $x=\sqrt{\alpha^2-1}$
these solutions read
\eqn\soldual{\eqalign{ 
W_i^{(1)}&=1_i+{\sqrt{\alpha^2-1}-1\over \alpha}\big( 
U_i^{(1)}+U_i^{(2)}+U_i^{(3)}\big)+U_i^{(4)}\cr
W_i^{(2)}&=1_i+{\sqrt{\alpha^2-1}-1\over \alpha}\big( 
U_i^{(1)}-U_i^{(2)}+U_i^{(3)}\big)+U_i^{(4)}\cr}}
where they not only differ by the sign in front of $U_i^{(2)}$
but also by $\alpha_2=\alpha_3=\sqrt{\alpha^2-1}$ in the case (1) and
$\alpha_2=\alpha_3=1$ in the case (2).

\subsec{The Loop Models}

In this section, we concentrate on the ``fundamental" solution
\funda\ found above, with 
$\alpha_2=\ldots=\alpha_{k-1}=\sqrt{\alpha_1^{2}-1}$, and
\eqn\wfund{\encadremath{
W_i(x) = 1_i +\sum_{q=1}^{k-1} {x^{q-1}\over
(\alpha_1^2-1)^{q-1\over 2}} {x-1\over \alpha_1} U_i^{(q)}
+{x^{k-1}\over (\alpha_1^2-1)^{{k\over 2}-1} } {x-1\over 
\alpha_1^2-1-x} U_i^{(k)} }}

Let us represent the projectors  
\ktyp\ as ``face operators" generalizing \sqconf, namely
\eqn\mulrep{ U_i^{(m)} = \figbox{5.cm}{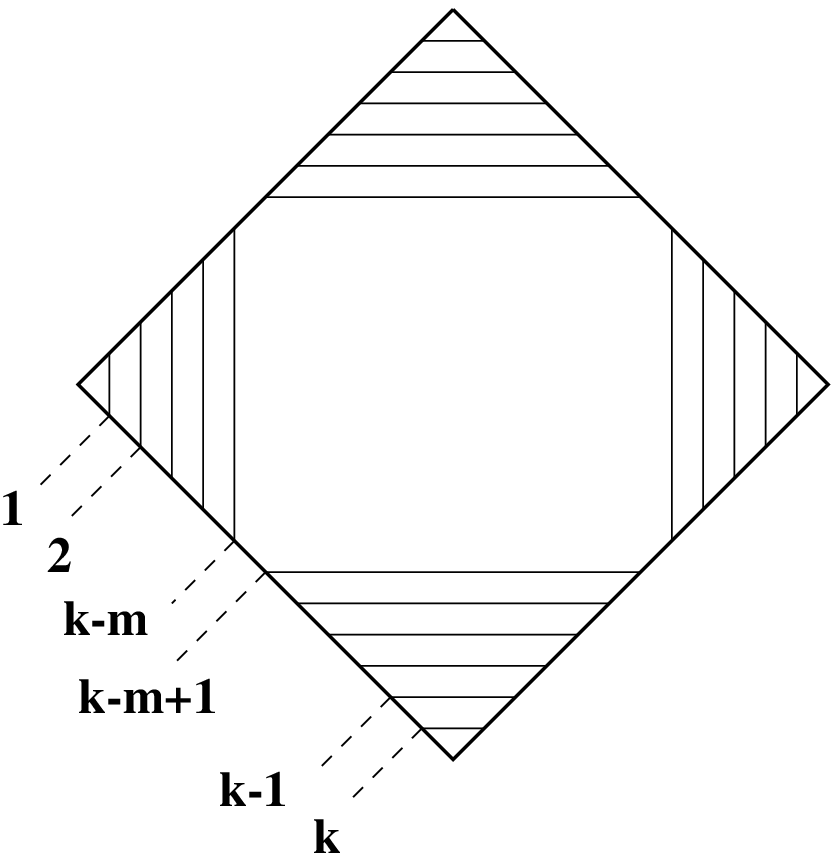} }
This gives rise to the following partition function
for the mutlicolored dense loop model, generalizing \pfuca, with
loops of colors $1,2,\ldots,k$:
\eqn\mucop{ Z = \sum_{{\rm face}\ {\rm configs}} 
\alpha_1^{L_1+L_k} 
(\alpha_1^{2}-1)^{{1\over 2}(L_2+\ldots+L_{k-1})} 
\prod_{m=1}^k a_m(x)^{N_m} }
where $L_i$ is the number of loops of color $i$, and $N_m$
the number of occurrences of the face corresponding to $U_i^{(m)}$.

As before, the models obeys a crossing symmetry relation, expressing
the covariance of the Boltzmann weights
under a rotation of $90^{\circ}$ of the faces
\eqn\covaq{ {\bar W}_i(x_*^2/x) =x_*^{k-2}
{x_*^2-x\over x^{k-1}(x-1)}\  W_i(x) }
with $x_*=\sqrt{\alpha_1^{2}-1}$, and where the bar stands for the 
rotation of $90^{\circ}$, with ${\bar U}_i^{(m)}=U_i^{(k-m)}$, for
$m=0,1,\ldots,k$ (we define $U^{(0)}=1$). The rotationally invariant
weights are $W_i(x_*)$. In the corresponding model, the loops
are weighted as in \mucop, whereas 
$a_1(x_*)=a_2(x_*)=\ldots=a_{k-1}(x_*)=(\sqrt{\alpha_1^{2}-1}-1)/\alpha_1$,
and $a_k(x_*)=1$.

\newsec{Discussion and Conclusion}

\subsec{Coloring Algebras}

The Fuss-Catalan algebras can also be viewed as subalgebras of
$TL_n(\alpha_1)\otimes TL_n(\alpha_2)\otimes\ldots\otimes
TL_n(\alpha_k)$ (with generators $u_i^{(m)}\in TL_{n}(\alpha_m)$), 
by expressing their generators as
\eqn\gefuka{ U_i^{(m)} = \left\{\matrix{
1\otimes 1\otimes \ldots\otimes 1\otimes 
u_i^{(k-m+1)}\otimes u_i^{(k-m+2)}\otimes \ldots\otimes u_i^{(k)}  
& \ {\rm if}\ i\ {\rm odd} \cr
u_i^{(1)}\otimes u_i^{(2)}\otimes \ldots\otimes u_i^{(m)}\otimes
1\otimes 1\otimes \ldots \ldots\otimes 1
& \ {\rm if}\ i\ {\rm even} \cr} \right. }
where there are exactly $k$ factors in both tensor products. For $k=2$
this reads
\eqn\keqto{ 1_i=1_a\otimes 1_b \qquad U_i^{(1)}=\left\{\matrix{
1_a\otimes U_b & \ {\rm if}\ i\ {\rm odd}\cr
U_a\otimes 1_b & \ {\rm if}\ i\ {\rm even}\cr}\right. \qquad
U_i^{(2)}= U_a\otimes U_b }
where the subscripts $a,b$ refer to the corresponding algebras $TL_n(\alpha),TL_n(\beta)$.

In principle, we could use this as a recipee for "coloring" any 
algebra. In particular, let us color the Hecke algebra $H_n(\beta)$,
by introducing generators $V_i^{(m)}$, in the same way as in \gefuka,
but using the generators of the Hecke algebras $H_n(\alpha_m)$
instead of the $u^{(m)}$.
Looking for a solution of the Yang-Baxter system 
\ybe\-\norma\-\imposi\ of the form
\eqn\eqyu{ W_i(x) = 1+\sum_{m=1}^k a_m(x) V_i^{(m)} }
we have found that there is no non-trivial
solution unless the initial algebras
actually obey the Temperley-Lieb relations \tla. Hence, the coloring 
process is particular to Temperley-Lieb, in the sense that, only in 
that case, we get new solutions to the Yang-Baxter equation.

\subsec{ADE Colored Models}

The Boltzmann weights \wfund\ of the colored loop models are related to 
Temperley-Lieb algebra generators through \gefuka. 
An explicit model is obtained by choosing a particular representation
of these Temperley-Lieb algebras, to build a representation
of the corresponding Fuss-Catalan algebra.
For instance, when $k=2$, using the $4\times 4$ representation
\sixv\ for all the Temperley-Lieb algebras $TL_n(\alpha_{m})$, $m=1,2$,
we get a $16\times 16$ matrix representation for $W_i(x)$, which reads
in $4\times 4$ block-form
\eqn\sixteen{\eqalign{
W_{\rm odd}&=\pmatrix{ 1+aU & 0 & 0 & 0 \cr
0 & 1+(a+bz)U & bU & 0\cr
0 & bU & 1+(a+b/z)U & 0 \cr
0 & 0 & 0 & 1+aU \cr} \cr
W_{\rm even}&=\pmatrix{ 1 & 0 & 0 & 0 \cr
0 & (1+az)1+bz U & a1+bU & 0\cr
0 & a1+bU & (1+a/z)1+b/z U & 0 \cr
0 & 0 & 0 & 1 \cr}\cr} }
with $U$ as in \sixv, and $1$ the $4\times 4$ identity matrix. 

\fig{The inhomogeneous 32 vertex model corresponding to the two-color
Fuss-Catalan solution to the Yang-Baxter equation.
We have represented in the first (resp. second) line the 
non-vanishing odd (resp. even) vertices,
together with their Boltzmann weights: $w_1=a$, $w_2=1+az$,
$w_3=1+az+bz^2$, $w_4=1+az+b$, $w_5=a+bz$, $w_6=bz$, $w_7=b$, 
with $a$ and $b$ as in \sola\ and \valb. To actually get all the vertices
(their total numbers are indicated in parentheses), we must apply the two
following transformations: (i) under $0 \leftrightarrow 3$, the Boltzmann 
weights are unchanged; (ii) under $1 \leftrightarrow 2$, we must also 
change $z \to 1/z$. Note that the second line of vertices is 
obtained from the first by interchanging the two spaces over 
which the operator $W$ acts, in agreement 
with \keqto.}{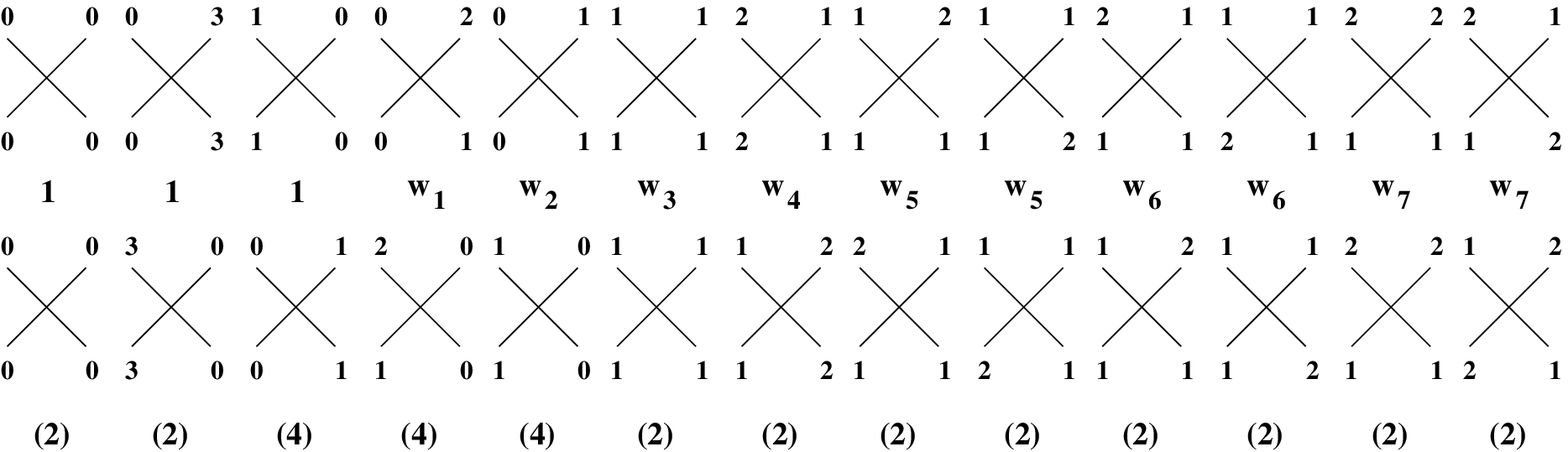}{12.cm}
\figlabel\vertex

We may therefore view the two-color model as an inhomogeneous 
$32$-Vertex model on the square lattice, the vertices being the
non-zero entries in \sixteen. These correspond naturally to
edge states taking four possible values $0,1,2,3$ (see Fig.\vertex).

Our result however is independent of the particular choices
of the representations 
of the Temperley-Lieb algebras. In particular, we could take 
different representations for the various factors $TL_n(\alpha_m)$.
For any connected non-oriented graph $\Gamma$ with adjacency matrix 
entries $G_{i,j}\in \{0,1\}$ and $G_{i,i}=0$, and each eigenvector
of $G$ with non-vanishing entries $S_i$ and eigenvalue $\beta$,
one has the following representation of $TL_n(\beta)$  
\eqn\reptla{1\left(\figbox{1.5cm}{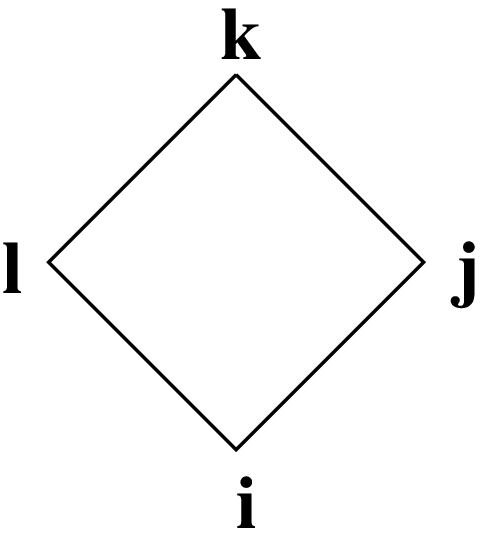} \right)
=\delta_{i,k},\qquad 
U\left(\figbox{1.5cm}{fijk.eps} \right)=
{\sqrt{S_i S_k}\over S_j }\delta_{j,l}}
where $i,j,k,l$ are now vertex variables on the square lattice, 
taking their values among the vertices of the target graph $\Gamma$,
and such that any two neighboring vertices of the lattice have 
values linked by an edge on $\Gamma$.
It has been shown that if $G=A,D,E$, one of the simply-laced
Dynkin diagrams, this representation leads through \mostl\ to the Boltzmann 
weights of some of the $A,D,E$ minimal models
(with largest eigenvalue $\beta=2\cos(\pi/m)$, $m=2,3,4,...$) 
\PAS\-\PEA\ 
(dilute vertex models must be 
considered as well to exhaust all the list, see \DIL\-\PEA), 
whereas the
extended Dynkin diagrams ${\hat A},{\hat D},{\hat E}$ (with
largest eigenvalue $\beta=2$) lead to the
free bosonic theory compactified on circles of specific radii. 

In the $k$-colored case, our ``fundamental" Boltzmann 
weights \wfund\ for arbitrary choices of $A,D,E$ representations in 
each factor give a number of new vertex models, with target space
a tensor product of $k$ simply-laced Dynkin diagrams. In the unitary case,
where all Boltzmann weights are positive, we must restrict the choice
of eigenvector to the Perron-Frobenius one, namely corresponding 
to the largest eigenvalue $\beta$. In that case, the target space
takes the form
$G_1\times G_2\times \ldots \times G_k$, where $G_i\in\{A,D,E\}$, and
where the maximum eigenvalue of $G_1,G_k$ is $\alpha_1=2\cos(\pi/m_1)$,
and that of $G_2,G_3,\ldots,G_{k-1}$ is 
$\sqrt{\alpha_1^{2}-1}=2\cos(\pi/m_2)$, 
where we have identified the Coxeter numbers 
$m_1$, $m_2$ of the Dynkin diagrams.  
This is actually only possible for the following values of $m_i$:
$(m_1,m_2)=(3,2);(6,4)$, 
hence only for
$(\alpha_1,\alpha_2)=(1,0);(\sqrt{3},\sqrt{2})$ 
(only the last case will have positive Boltzmann weights, 
when \positgen\ holds).
This leaves us with only the Dynkin diagrams of
$A_3$, $A_5$ and $D_4$, with respective Coxeter number
$4$, $6$ and $6$.

Even in general, the condition that both $\alpha_1$ and 
$\sqrt{\alpha_1^{2}-1}$ be eigenvalues of Dynkin diagrams is very
restrictive, and we are left only with 
$(\alpha_1,\alpha_2)=(1,0);(\sqrt{3},\sqrt{2})$.
These can appear as (non-necessarily Perron-Frobenius)
eigenvalues of a number of Dynkin diagrams.

We can also consider the case where we take ADE representations for say
$TL_n(\alpha_1)$ and $TL_n(\alpha_k)$, and a 6 vertex representation
for the other factors $TL_n(\alpha_i)$, $i=2,3,...,k-1$. In that case,
any ADE is allowed, and if $\alpha_1=2\cos(\pi/m)$, then
$\alpha_2=\sqrt{4\cos^2(\pi/m)-1}$ for $m=2,3,5,6,...$, where again
we have excluded $m=4$ as it leads to $\mu=1$. Conversely, we may take
any ADE representation for $TL_n(\alpha_i)$, $i=2,3,...,k-1$, and 
a 6 vertex representation for $TL_n(\alpha_1)$ and $TL_n(\alpha_k)$.
In the latter case, we have $\alpha_2=2\cos(\pi/m)$, and
$\alpha_1=\sqrt{4\cos^2(\pi/m)+1}$, $m=3,4,5,...$.

Finally, we could also take ${\hat A}{\hat D}{\hat E}$ representations
say for $TL_n(\alpha_1)$ and $TL_n(\alpha_k)$, 
when $\alpha_1=\alpha_k=2$, and then either $A_5$ or $D_6$ 
representations for $TL_n(\alpha_m)$, $m=2,3,...,k-1$, 
with all $\alpha_m=\sqrt{3}$. Conversely, we may take 
${\hat A}{\hat D}{\hat E}$ representations for 
$TL_n(\alpha_m)$, $m=2,3,...,k-1$, with all $\alpha_m=2$, and a
6 vertex representation for $TL_n(\alpha_1)$ and $TL_n(\alpha_k)$,
with $\alpha_1=\alpha_k=\sqrt{5}$.

Allowing for excited solutions with some negative $r_i$'s, this
gives the possibility of choosing an $A_2$ representation for the
$TL_n(\alpha_i)$ factors with $\alpha_i=1=2\cos(\pi/3)$ 
(namely such that $r_i=-r_{i-1}$).

\subsec{Rational Limit}

The Boltzmann weights \wfund\ admit a rational limit, obtained by 
setting $x=e^{\epsilon u}$ and $\alpha_1^{2}-1=e^{\epsilon}$, and
letting $\epsilon \to 0$.
We immediately get
\eqn\immedi{ W_i^{rat}(u) = 1_i + { u \over 1- u} U_i^{(k)} }
But, thanks to \fucalge, the $U_i^{(k)}$ generate a Temperley-Lieb algebra
with parameter $\beta=\rho_k=2$ here, hence \immedi\ is the standard
rational solution to \ybe. The same limit is also obtained from the
Temperley-Lieb vertex model \mostl\-\tla, for $x=e^{\epsilon u}$,
$z=e^{\epsilon/2}$, and $\epsilon\to 0$. 

\subsec{Open Questions}

The first question to be answered is whether the solutions \wfund\
and their excitations are new. To our knowledge, they are. 
We can compare them with the Boltzmann weights of
the fused 6 Vertex model, which read (in the simplest ``2-fused" case of the
19 Vertex model)
\eqn\wnineteen{ W_i(x) =  1_i + {x-1\over z-x/z} \left(P_i^{(1)}+
{z x-1\over z^{2}-x/z}  P_i^{(2)}\right) }
where the $P$'s are unnormalized projectors. We see the emergence of 
new poles in $x$, which does not occur in our case.
Assuming the models are new, it would also be interesting to
find the structure of their fusions, in algebraic terms.

As the model \wfund\ is integrable, one should be able to solve it
by Bethe Ansatz techniques. The thermodynamic limit will presumably 
display some interesting phase diagram, of which the second order
critical points should correspond to conformal theories. This
program will be carried out elsewhere.
It would also be desirable to have an elliptic version of the solutions
\wfund, generalizing Baxter's 8 vertex model Boltzmann weights.

The 6 vertex model also describes the XXZ quantum spin chain, closely
related to the affine $sl_2$ quantum group. One could wonder whether
some ``colored" quantum spin chains actually correspond to our models.

Finally, let us recall the connection between Temperley-Lieb algebra 
and multi-component meanders, established in \TLM. 
A multi-component meander is a configuration of 
closed non-intersecting loops crossing a line through a given number 
of points, and may also be viewed as a compactly folded configuration 
of several possibly interlocked polymer chains. 
The Fuss-Catalan algebras allow for the definition of 
multi-colored meanders, with very analogous properties \MDET.

\bigskip

\noindent{\bf Acknowledgements}

We thank A. Ram for pointing out ref.\FC\ to us,
and P. Pearce for interesting discussions. This work
was partly supported by the NSF grant PHY-9722060. 

\listrefs 
\end